\documentclass[aps, twocolumn, superscriptaddress, showpacs, floatfix]{revtex4-1}
\usepackage{amsmath}
\usepackage{bbm}
\usepackage{graphicx}
\usepackage{float}
\usepackage{multirow}
\usepackage[usenames,dvipsnames]{xcolor}
\usepackage[colorlinks=true,linkcolor=blue,citecolor=blue,urlcolor=blue]{hyperref}

\newcommand{\EF}{E_{\text{F}}}
\newcommand{\ud}{\text{d}}
\newcommand{\iu}{\text{i}}
\newcommand{\IM}{\text{Im}}
\newcommand{\RE}{\text{Re}}
\newcommand{\HH}{\mathcal{H}}

\begin{document}

\title{Strong correlation effects in theoretical STM studies of magnetic adatoms}
\author{Hung T. Dang}
\affiliation{Institute for Theoretical Solid State Physics, JARA-FIT and JARA-HPC, RWTH Aachen University, 52056 Aachen, Germany}
\author{Manuel dos Santos Dias}
\affiliation{Peter Gr\"{u}nberg Institut and Institute for Advanced Simulation, Forschungszentrum J\"{u}lich \& JARA, D-52425 J\"{u}lich, Germany}
\author{Ansgar Liebsch}
\affiliation{Peter Gr\"{u}nberg Institut and Institute for Advanced Simulation, Forschungszentrum J\"{u}lich \& JARA, D-52425 J\"{u}lich, Germany}
\author{Samir Lounis}
\affiliation{Peter Gr\"{u}nberg Institut and Institute for Advanced Simulation, Forschungszentrum J\"{u}lich \& JARA, D-52425 J\"{u}lich, Germany}

\begin{abstract}
We present a theoretical study for the scanning tunneling microscopy (STM) spectra of surface-supported magnetic nanostructures, incorporating strong correlation effects. As concrete examples, we study Co and Mn adatoms on the Cu(111) surface, which are expected to represent the opposite limits of Kondo physics and local moment behavior, using a combination of density functional theory and both quantum Monte Carlo and exact diagonalization impurity solvers. We examine in detail the effects of temperature $T$, correlation strength $U$, and impurity $d$ electron occupancy $N_d$ on the local density of states. We also study the effective coherence energy scale, i.e., the Kondo temperature $T_K$, which can be extracted from the STM spectra. Theoretical STM spectra are computed as a function of STM tip position relative to each adatom. Because of the multi-orbital nature of the adatoms, the STM spectra are shown to consist of a complicated superposition of orbital contributions, with different orbital symmetries, self-energies and Kondo temperatures. For a Mn adatom, which is close to half-filling, the STM spectra are featureless near the Fermi level. On the other hand, the quasiparticle peak for a Co adatom gives rise to strongly position-dependent Fano line-shapes.
\end{abstract}

\maketitle

\section{Introduction\label{sec:intro}}
Over the last thirty years, scanning tunneling microscopy (STM) has opened the window into the nanoscale realm. Its versatility ranges from topographic characterization to spatially resolved spectrum acquisition. More recently, it has enabled the mapping of magnetic properties down to the single atom level, allowing to gain more insights into the fundamental problem of how a single magnetic adatom behaves when placed on a surface and coupled to itinerant electrons~\cite{Wiesendanger2009}.

Besides probing the surface electronic structure around the Fermi level $\EF$, low-energy excitations can also be detected if the bias voltage applied to the STM tip is greater than the corresponding excitation energy. This has been utilized to characterize phonon modes in adsorbed molecules~\cite{Lorente2005}, spin excitations in adatoms and small clusters~\cite{Heinrich04,Hirjibehedin2007,Khajetoorians2011a,Khajetoorians2013}, and many-body effects, especially Kondo resonances (see, e.g., Refs.~\cite{Li98,Madhavan98,Knorr2002,Nagaoka2002,Otte2008,Ternes2009,Pruser2011,Prueser2012,Pruser2014,VonBergmann14,Ternes2015}).

Magnetic adatoms on surfaces can exhibit various phenomena, from quantum spins on insulating surfaces~\cite{Hirjibehedin2007,Loth2012,Rau2014} to a more itinerant behavior on metallic surfaces~\cite{Chilian2011,Khajetoorians2011a}. The determining factor is the competition between the local Coulomb interaction among the $d$ electrons of the adatom and the coupling to the surface electrons, which tends to screen the interaction~\cite{book:Hewson97}.

The theoretical description of STM begins by computing the electronic structure of the surface in the presence of the adatom of interest. Density functional theory (DFT) is the standard approach, together with a model for the interaction of the tip with the surface~\cite{Wortmann2001,Palotas2012}. When the behavior of the adatom is dominated by strong local correlations, this mean-field description fails. A combination of DFT with many-body methods is then the appropriate route, as shown by previous studies~\cite{Carter2008,Jacob09,Surer12,Baruselli12,Gardonio13,Jacob15}. The final step is to evaluate the modification of the STM spectra caused by the effects of strong correlations at the adatom. Of particular interest is the issue of how these multi-orbital correlations influence the STM line shape as a function of tip position relative to the adsorbed atom. In previous studies, this step was usually not taken into consideration, with few exceptions~\cite{Ujsaghy2000,Plihal2001,Schweflinghaus2014,Frank2015,Baruselli2015}.
Thus, the goal of the paper is to present a complete description of STM spectra of strongly correlated adatoms, using state-of-the-art single-particle and many-body methods.

We choose two $3d$ transition metal adatoms which are expected to represent qualitatively different behavior: Co, a prototypical Kondo system~\cite{Knorr2002}, and Mn, close to the local moment limit~\cite{Gardonio13} (half-filled $d$ shell), which are deposited on the Cu(111) surface. The Korringa-Kohn-Rostoker (KKR) method~\cite{Papanikolaou2002} is employed for the description of the electronic structure of the surface in the presence of the adatom. Combining KKR with a continuous time quantum Monte Carlo (CT-QMC) impurity solver~\cite{Werner06,Gull11} allows us to properly treat strong correlations in the $d$ shell of the adatoms. In addition, we have used exact diagonalization (ED)~\cite{Caffarel1994,Liebsch2012} impurity solver in order to explore temperatures below those that are presently accessible within CT-QMC.

The main result of the approach outlined above is that the STM spectra comprise complex superpositions of five orbital contributions, each of which is associated with different single-particle hybridization functions and self-energies. Accordingly, depending on the vertical and lateral positions of the tip with respect to the adatom, the STM spectra exhibit strongly varying Fano profiles that cannot be simply represented in terms of single-orbital models. These results suggest the possibility of extracting from experiment the different orbital-dependent Kondo temperatures based on the anisotropic environment of the atom adsorbed on the metallic surface. 
               
The paper is organized as follows. Section~\ref{sec:model_method} introduces the model and the theoretical approaches to study the adatom supported on the copper surface. Section~\ref{sec:kkr} presents the electronic structure obtained from DFT using the KKR method and the orbital-dependent hybridization functions which form the input in the subsequent many-body calculations. Section~\ref{sec:corr} explores the correlation effects in the adatom $d$ shell and their dependence on different parameters, such as temperature, interaction strength, and electron filling. In Sec.~\ref{sec:stm}, we present the calculated STM spectra. In particular, we show how strong correlations in the $d$ shell impact the electronic structure of the surface away from the adatom. Section~\ref{sec:conclusions} provides the conclusion. Two appendices describe in more detail several technical aspects: the main features of the exact diagonalization impurity solver (Appendix~\ref{app:exactdiagonalization}), and a comparison of orbital occupancy and self-energy results with those obtained for Lorentzian hybridization functions (Appendix~\ref{app:lorentzian}).

\section{Theory and methods\label{sec:model_method}}

\begin{figure}[t]
 \includegraphics[width=0.8\columnwidth]{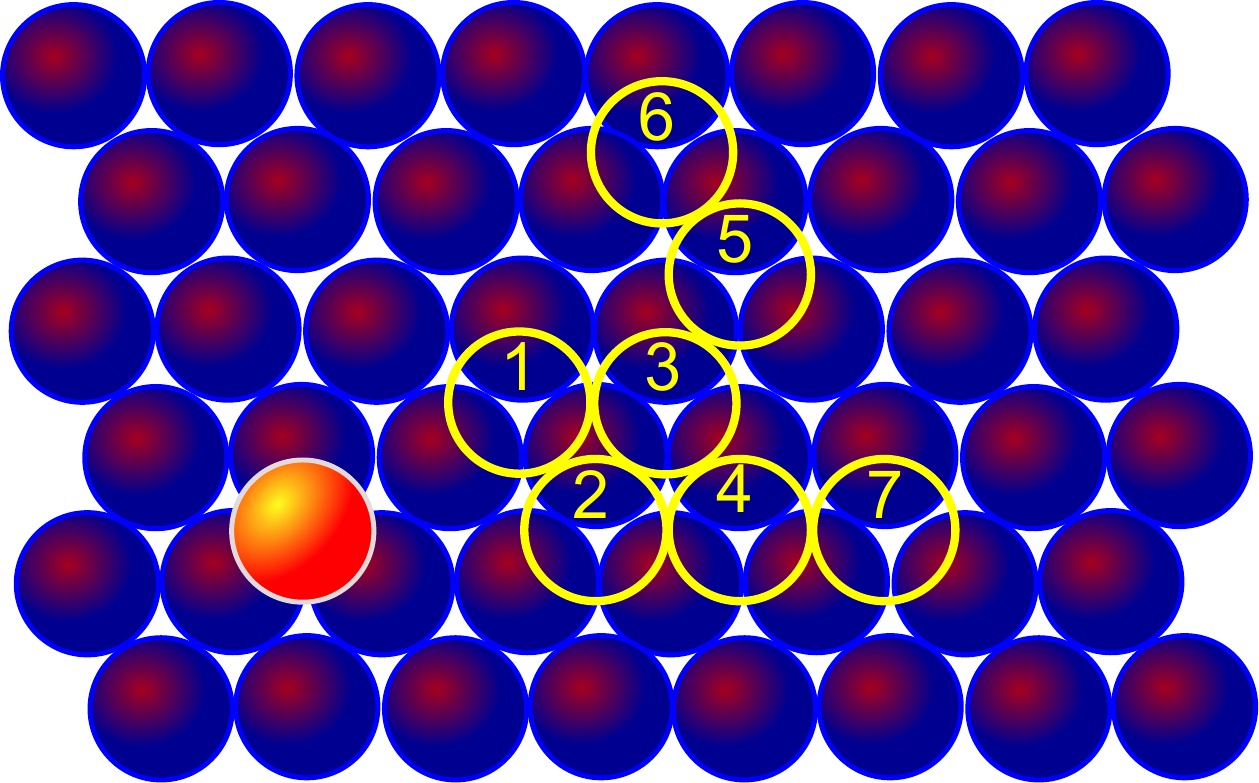}
\caption{\label{fig:impurity_cartoon}(Color online) Schematic plot of an adatom (Mn or Co - light red ball) placed on the $(111)$ surface of copper (dark balls). The viewpoint is along the surface direction of copper. Light (yellow online) circles mark possible positions of the STM tip investigated in this paper, which are at the same level as the adatom.}
\end{figure}

The investigated system is sketched in Fig.~\ref{fig:impurity_cartoon}. A transition metal adatom (Mn or Co) is adsorbed on a hollow site of the Cu(111) surface \cite{Manoharan00,Madhavan98}, supported by three Cu surface atoms in a triangular arrangement. In correspondence with experiment, the STM tip is allowed to move parallel to surface plane as well as vertically above the adatom. 

The most elementary description of an STM measurement is based on a simplified Tersoff-Hamann model~\cite{Tersoff1983,Wortmann2001}. In this scheme, the differential conductance in the tunneling regime, $\ud I/\ud V$, is assumed to be proportional to the local DOS of the substrate, $\rho_s$, measured around the Fermi energy, $\EF$, by varying the bias voltage, $V$ (with $e$ the electron charge), and spatially integrated over a certain region around the tip position, $\vec{R}_t$:
\begin{equation}\label{eq:thmodel}
  \frac{\ud I}{\ud V} \propto \rho_s(\EF+eV;\vec{R}_t). 
\end{equation}
For simplicity, the energy dependence of the tip DOS and of the tunneling matrix between the tip and the sample are neglected.

The key quantity, the local density of states (DOS) at the tip location, reads
\begin{equation}\label{eq:rhos}
  \rho_s(\EF+eV;\vec{R}_t) = -\frac{1}{\pi}\,\IM\!\int_{V_t}\!\!\!\ud\vec{r}\;G_s(\vec{r}\,,\vec{r}\,;\EF+eV).
\end{equation}
The single-particle Green's function describing the electronic structure of the surface in the presence of the adatom can be obtained via a Dyson equation
\begin{align}
  &\hspace{-0.3em}G_s(\vec{r}\,,\vec{r}\,';E) = G^0_s(\vec{r}\,,\vec{r}\,';E) \nonumber\\
  &\hspace{-0.3em}+ \!\int\!\!\ud\vec{r}_1 \!\!\int\!\!\ud\vec{r}_2\,G^0_s(\vec{r}\,,\vec{r}_1;E) \Sigma_a(\vec{r}_1,\vec{r}_2;E) G_s(\vec{r}_2,\vec{r}\,';E) \;.
\end{align}
Here $G^0_s$ is the Green's function for the pristine surface, possessing two-dimensional translational invariance. It is obtained via DFT. The presence of the adatom acts as a local perturbation to the surface. Its effects on the electronic structure are described by a self-energy
\begin{equation}
  \Sigma_a(\vec{r}\,,\vec{r}\,';E) = \Delta V_{\text{KS}}(\vec{r}\,)\,\delta(\vec{r}\,-\vec{r}\,') + \Sigma_c(\vec{r}\,,\vec{r}\,';E) \;.
\end{equation}
From the DFT point of view, the presence of the adatom results in a change of the surface Kohn-Sham potential, $\Delta V_{\text{KS}}$, in a finite region surrounding the adatom. The effect of correlations within the $d$ shell of the adatom gives the self-energy correction $\Sigma_c$.

\subsection{DFT Green's function formalism}

The electronic structure of the surface is obtained using the KKR Green's function method~\cite{Papanikolaou2002} in the atomic sphere approximation (ASA) considering the full charge density. The KKR Green's function is given by
\begin{align}
  G_{ij}(\vec{r}\,,\vec{r}\,'&;E) = \sum_{LL'} Y_L(\hat{r}) \Big[\delta_{ij}\delta_{LL'} R_{i\ell}(r_<;E) H_{i\ell}(r_>;E) \nonumber\\
  &+ R_{i\ell}(r;E) G_{iL,jL'}(E) R_{j\ell'}(r';E)\Big]Y_{L'}(\hat{r}'),
\end{align}
where $i$ and $j$ label lattice positions at which atomic spheres are centered.
Here $\vec{r}\, = |\vec{r}|\frac{\vec{r}}{|\vec{r}|} = r\,\hat{r}$, $r_< = \min\{r,r'\}$, $r_> = \max\{r,r'\}$, $L = (\ell,m)$ is a combined angular momentum index, $Y_L(\hat{r})$ are real spherical harmonics and $R_{i\ell}(r;\epsilon)$ and $H_{i\ell}(r;\epsilon)$ are regular and irregular scattering solutions to the Kohn-Sham potential in atomic sphere $i$ for a given energy $\epsilon$. The elements of the structural Green's function, $G_{iL,jL'}(\epsilon)$, contain the multiple scattering contributions.

The calculations use the local density approximation (LDA) as parametrized by Vosko, Wilk and Nusair~\cite{Vosko1980}. The Cu(111) surface is represented by a slab of 22 Cu layers, extended by two vacuum regions equivalent to four more layers above and below the slab. Instead of a supercell approach, the change in the electronic structure caused by the presence of the adatom is computed directly in real space, using an embedding procedure. The adatom is relaxed towards the surface by 10\% of the ideal Cu--Cu interlayer spacing; since it is negligible, no relaxation of the Cu surface layers was considered. All calculations are non-spin-polarized; the spin physics will be described within the correlated model.

In the ASA, space is divided into spheres surrounding lattice points; in the present case this means all layers are partitioned in this way, including the vacuum layers. The adatom replaces one of the vacuum spheres on the layer above the Cu surface layer. The ASA spheres are taken as the averaging volume $V_t$ in Eq.~\eqref{eq:rhos}, allowing a discrete sampling of the DOS away from the adatom, parallel or normal to the surface.

We define projector operators for the adatom $d$ orbitals as follows:
\begin{equation}\label{eq:projector}
  P_m(\vec{r}\,,\vec{r}\,') = Y_2^m(\hat{r})\phi(r)\,\phi(r')Y_2^m(\hat{r}') \,,\; \{\vec{r}\,,\vec{r}\,'\} \in V_a \,.
\end{equation}
The angular dependence of the orbitals is represented by real spherical harmonics $Y_2^m(\hat{r})$, $m = \{xy$, $yz$, $3z^2-r^2$, $xz$, $x^2-y^2\}$, and $\phi(r)$ are real radial functions built by normalizing regular scattering solutions computed at the Fermi energy to unity inside the atomic sphere surrounding the adatom, $V_a$. These projectors are used to define the correlated subspace to fully treat the many-body physics, as explained in the next section.

\subsection{Treatment of strong correlations}

The Hilbert space spanned by the $d$ orbitals of the adatom is taken to be the correlated subspace. Electrons from the rest of the system can hop in and out of these orbitals, leading to a multi-orbital Anderson impurity model
\begin{align}\label{eq:impurity_model}
  \HH &= \sum_{m\sigma} \epsilon_{m}\,d^\dagger_{m\sigma}d_{m\sigma} + \sum_{\alpha\beta\gamma\delta} U_{\alpha\beta\gamma\delta}\,d^\dagger_{\alpha}d^\dagger_{\beta}d_{\gamma}d_{\delta} \nonumber\\
  &+ \sum_{p\sigma} \epsilon_{p}\,c^\dagger_{p\sigma}c_{p\sigma} + \sum_{mp\sigma} V_{mp}\,d^\dagger_{m\sigma}c_{p\sigma} + \text{h.c.}
\end{align}
The Hamiltonian is expressed in terms of creation and annihilation operators for localized $d$-electrons, $d^\dagger_{m\sigma}$ and $d_{m\sigma}$, for orbital $m$ and spin $\sigma$ or combined indices $\alpha=(m,\sigma)$, and the conduction electrons of the surface, $c^\dagger_{p\sigma}$ and $c_{p\sigma}$, which are given some generic state label $p$ with spin index. We drop the spin index $\sigma$ from this point onwards, because there is no spin symmetry breaking in the calculations.

The interaction term 
$\HH_{\text{int}}=\sum_{\alpha\beta\gamma\delta} U_{\alpha\beta\gamma\delta}\,d^\dagger_{\alpha}d^\dagger_{\beta}d_{\gamma}d_{\delta}$ 
is formulated phenomenologically based on Slater's integrals $F_0$, $F_2$ and $F_4$~\cite{Slater60}, which is a rotationally invariant form. It is parametrized by $U = F_0 = 4$ or $5$~eV and $J=(F_2+F_4)/14 = 0.9$~eV, similar to the values used in earlier studies~\cite{Surer12,Gardonio13}. The ratio $F_4/F_2 = 0.63$.

The remaining quantities appearing in Eq.~\eqref{eq:impurity_model}, the energy levels $\epsilon_{m,p}$ and hopping matrix elements $V_{mp}$, are constructed from DFT Green's function. The interacting Green's function for the localized orbitals is written as 
\begin{equation}\label{eq:gfmatsu}
  G_{m}(\iu\omega_n) = \frac{1}{\iu\omega_n + \mu - \epsilon_m - \Delta_m(\iu\omega_n) - \Sigma_{m}(\iu\omega_n)}\;\;, 
\end{equation}
where $\omega_n$ is a fermionic Matsubara frequency and $\mu$ the chemical potential. The non-interacting Green's function $G^0_m(\iu\omega_n)$ ($\Sigma_m(\iu\omega) = 0$) is obtained directly from the DFT electronic structure via the projection operators defined in Eq.~\eqref{eq:projector}. The impurity energy level $\epsilon_m$ is derived from the tail of this Green's function. The single-particle coupling between the impurity and the rest of the system is expressed in terms of the hybridization function
\begin{equation}\label{eq:hybridization}
  \Delta_m(\iu\omega_n) = \sum_{p}\frac{|V_{mp}|^2}{\iu\omega_n - \epsilon_p} = \Delta_m'(\iu\omega_n) + \iu\Delta_m''(\iu\omega_n)\;\;,
\end{equation}
where the real and imaginary parts are denoted by single and double primes, respectively. 

As will be discussed later, the DFT calculations show that all five $d$ orbitals are located near the Fermi level and are partially filled. Thus, for a dynamical treatment, it is necessary to consider them simultaneously. To solve the highly non-trivial five-orbital Anderson impurity model, we use mainly the hybridization expansion version (CT-HYB) of the continuous-time quantum Monte Carlo impurity solver~\cite{Werner06,Gull11} implemented in the TRIQS package~\cite{Parcollet15,Seth15}. In addition, the exact diagonalization (ED) method \cite{Caffarel1994,Liebsch2012} is used. Both impurity solvers may be viewed as tools for evaluating the local adatom self-energy
$\Sigma_m(\iu\omega) = \Sigma_m'(\iu\omega) + \iu\Sigma_m''(\iu\omega)$, where QMC has computational advantages at higher temperatures, while ED is useful to access somewhat lower temperatures than those presently available within QMC. Nonetheless, because of finite-size limitations, ED cannot reach arbitrarily low temperatures either. This problem is discussed in greater detail in Appendix~\ref{app:exactdiagonalization}. At temperatures where both schemes overlap, the computed self-energies were found to be in nearly quantitative  agreement.
         
As is well known, there is a double-counting problem which occurs also in DFT+$U$, DFT+DMFT, etc. (see e.g. Refs.~\onlinecite{Anisimov91,Czyzyk94,Nekrasov13,Wang12,Dang14,Haule2014}). The correlation effects on the $d$ orbitals of the adatom are treated twice, once via the LDA exchange-correlation potential, and again via the solution of the Anderson impurity model. To compensate, a double counting correction is introduced to subtract the correlation energy contained in the \textit{ab initio} calculation. However, the exact form of this correction is not well established. In the present systems, this correction can be conveniently embedded into the chemical potential which changes monotonically with the occupancy of the correlated $d$ orbitals $N_d$~\cite{Wang12,Dang14}. Here we consider a wide range of values for $N_d$, which include the physical $d$ electron occupancies.

\subsection{Analytical continuation\label{subsec:acont}}

The Green's function [Eq.~\eqref{eq:gfmatsu}] and the self-energy obtained via the impurity solver are evaluated at Matsubara frequencies. To relate these functions to experimental STM data, they need to be expressed at real frequencies, thus requiring an analytical continuation procedure. In the case of CT-QMC, initially we use the maximum entropy (MaxEnt) method~\cite{Jarrell96}, which allows us to access the spectra in a wide range of energies. 

However, to access the narrow energy window around the Fermi energy ($\EF \pm 0.1$~eV) relevant for the STM spectra, MaxEnt becomes unstable unless the QMC simulation is carried out with better statistics than presently available. Therefore, we consider an alternative approach by fitting the self-energy to a rational function
\begin{equation}\label{eq:ratfit}
  \Sigma_m(\iu\omega_n) = \Sigma_m(\iu\infty) + \frac{\sum_{p=0}^{N-1} a_p (\iu\omega_n)^p}{1 + \sum_{q=1}^{N} b_q (\iu\omega_n)^q},
\end{equation}
where $\Sigma_m(\iu\infty)$ is the Hartree shift of self-energy, and the parameters $a_p$ and $b_q$ can be obtained via linear least-squares fitting to the QMC data (the first $6N$ Matsubara frequencies are used, where $N=3$ is a good compromise).

\vspace{2em}

\subsection{Coherence scale $T_K$}

In the impurity model, there exists an important coherence scale, the Kondo temperature $T_K$, below which the local moment in the impurity is strongly screened by the electrons nearby. Equivalently, it marks the onset of the Fermi liquid behavior in the impurity \cite{book:Hewson97}. As a result, $T_K$ is proportional to the spectral weight at the Fermi level, characterized by the quasi-particle 
renormalization factor $Z$. Since the self-energy is diagonal in orbital space, the components of $Z$ are given by 
\begin{equation}\label{eq:renorm_factor}
  Z_m^{-1} = 1 - \left.\frac{\partial\Sigma_m'}{\partial\omega}\right|_{\omega=0} \!\!= 1 - \left.\frac{\partial\Sigma_m''}{\partial\omega_n}\right|_{\omega_n=0}.
\end{equation}
Thus, because of the analytic properties of the self-energy, $Z_m$ can be evaluated directly on the Matsubara axis. We then estimate the derivative $\partial \Sigma_{m}''/\partial \omega_n|_{\omega_n \to 0}$ by using fifth-order polynomial fitting of $\Sigma_{m}(i\omega_n)$ for the first six Matsubara frequencies together with an extra point $(0, 0)$ (in a Fermi liquid, $\Sigma''(i\omega_n)\to 0$ when $\omega_n\to 0$, i.e. $T\to 0$). The additional point is important to stabilize the fitting procedure, giving consistent results for various cases when going from Mn to Co adatoms.

As from previous works, we use  the following formula to estimate the orbital dependent Kondo temperature \cite{book:Hewson97,Surer12,Hewson05}:
\begin{equation}\label{eq:kondoT}
  T_{K,m} = \frac{\pi}{4} Z_{m} \Gamma_{m},
\end{equation}
where $\Gamma_{m} = -\Delta_m''(\omega\approx0)$ the imaginary part of hybridization function at Fermi level (see Sec.~\ref{sec:kkr}).

\section{DFT calculations}\label{sec:kkr}

We first discuss the results related to the surface electronic structure in the presence of adatoms using band structure calculations and extract the relevant quantities to construct the multi-orbital Anderson impurity model [Eq.~\eqref{eq:impurity_model}]. We note that, because the adatom is placed on the hollow site of the Cu (111) surface, the point group symmetry is $C_{\text{3v}}$. Thus the $d$ orbitals are split into three groups of degenerate orbitals: $E_2 = \{x^2-y^2, xy\}$, $E_1 = \{xz,yz\}$ and $A_1 = \{3z^2-r^2\}$ \cite{Surer12}. These notations will be employed throughout the paper. 

\begin{figure}[t]
 \centering
 \includegraphics[width = \columnwidth]{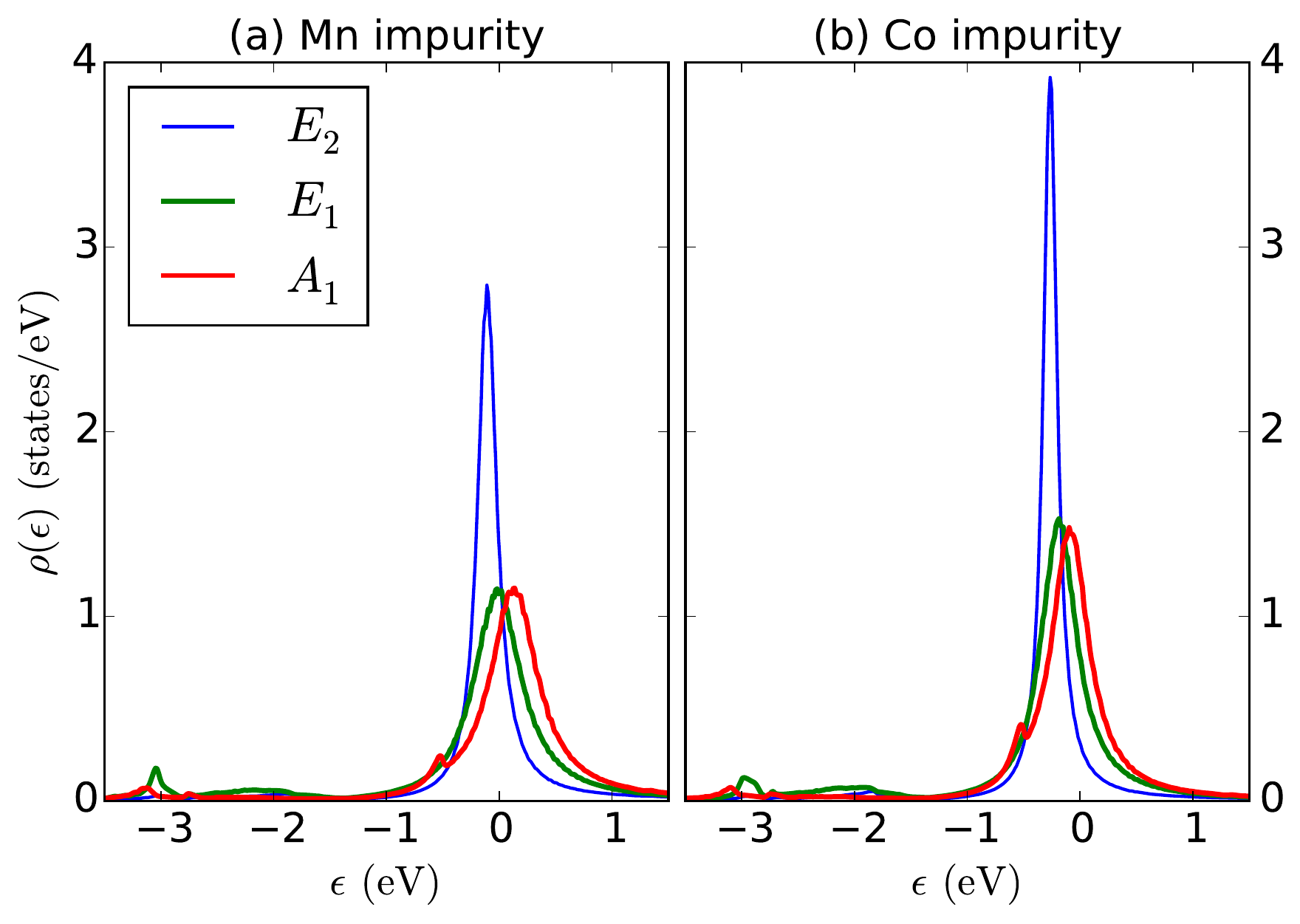}
\caption{\label{fig:kkr_dos}(Color online) Local DOS $\rho^0_m(\epsilon)$ from LDA calculations for Mn (a) and Co (b) impurities on the $(111)$ surface of copper. The energy zero corresponds to the Fermi energy of the Cu(111) substrate.}
\end{figure}

Figure~\ref{fig:kkr_dos} shows the local DOS of the adatoms obtained from the LDA calculations. The orbital decomposition of the DOS is achieved through the use of the projectors defined in Eq.~\eqref{eq:projector}. Mn and Co adatoms exhibit rather similar features: all of the $d$ orbitals are located near the Fermi energy and are partially filled. The $d$ orbitals of Co are lower in energy than those of Mn as the $d$ occupation is larger for Co than for Mn. The small features at $\epsilon \approx -3$~eV are associated with the hybridization with Cu $d$ states. The small peak near $\epsilon \approx -0.5$~eV present only for the $A_1$ orbital is assigned to the coupling between this orbital and the Cu(111) surface state, which starts near this energy~\cite{Limot2005,Lounis2006}.

To understand the main peak near the Fermi level, we start from Eq.~\eqref{eq:gfmatsu}. The non-interacting DOS for orbital $m$ is given by
\begin{equation}
  \rho_m^0(\omega) = -\frac{1}{\pi}\frac{\Delta_m''(\omega)}{\big(\omega + \mu - \epsilon_m - \Delta_m'(\omega)\big)^2 + \big(\Delta_m''(\omega)\big)^2} \;\;.
\end{equation}
At low frequencies, $\Delta_m'(\omega)$ and $\Delta_m''(\omega)$ are approximately constant, implying a Lorentzian form of the local DOS:
\begin{equation}\label{eq:lorentzian}
  \rho_m^0(\omega) \approx \frac{1}{\pi}\frac{\Gamma_m}{\big(\omega + \mu - \epsilon_m\big)^2 + \big(\Gamma_m\big)^2} ,
\end{equation}
where $\Delta_m'(\omega\approx0)$ is absorbed into the chemical potential $\mu$, while $\Gamma_m=-\Delta_m''(\omega\approx0)$ defines the Lorentzian line width.

There is uncertainty in specifying the $d$ occupancy $N_d$ of the adatom. The orbital-projected DOS is unnormalized, i.e. $\int\!\ud\epsilon\,\rho_m^0(\epsilon) < 1$. This stems from the projector approximation, which assumes the $d$ orbitals are completely localized on the adatom atomic sphere. By normalizing the integrated DOS to unity, we obtain the occupancies $5.75$ for Mn and $7.86$ for Co; these values are used as DFT $N_d$ values for reference throughout the paper.

\subsection{Hybridization function}

\begin{figure}[t]
\centering
 \includegraphics[width=\columnwidth]{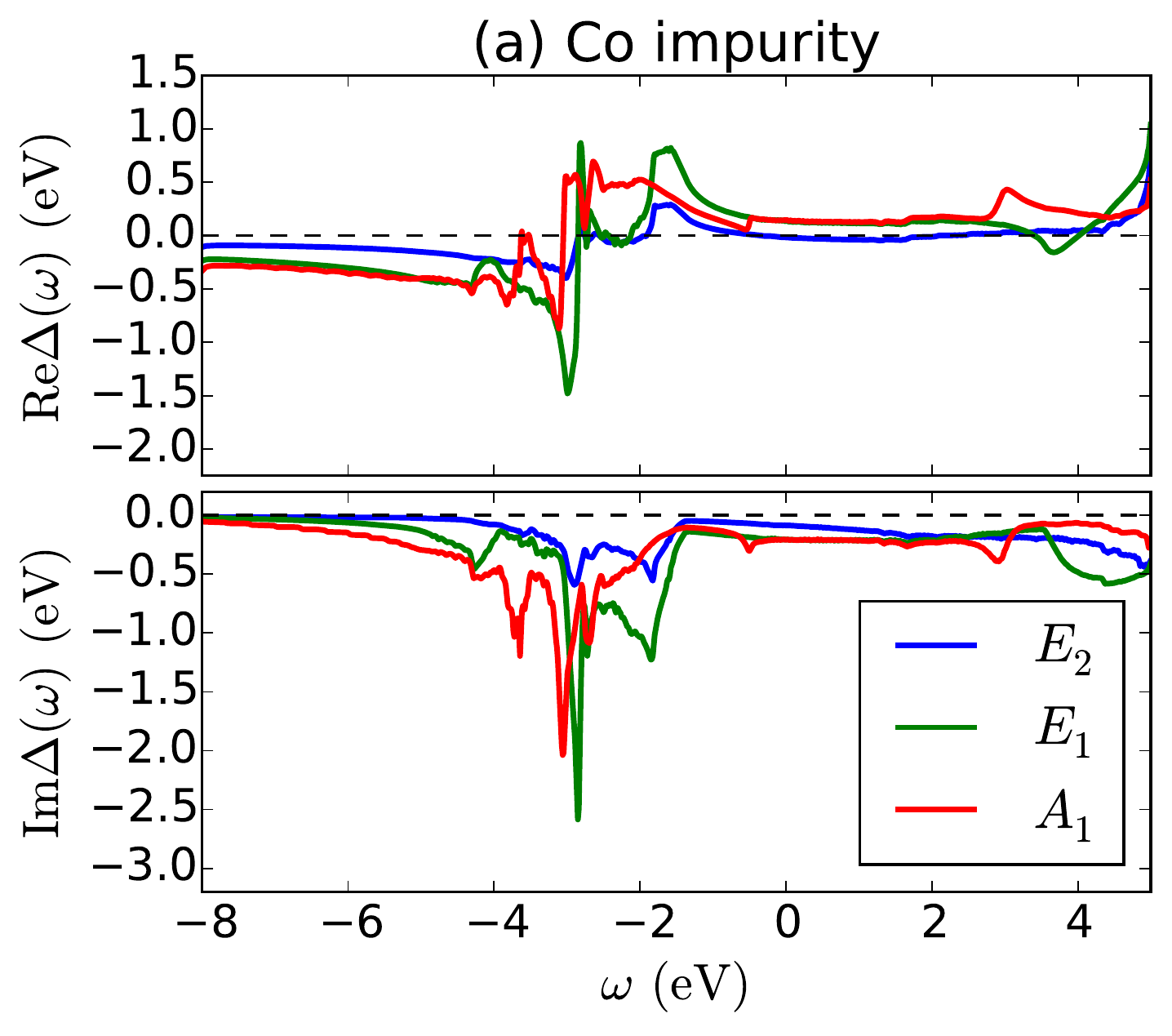}\\
 \includegraphics[width=\columnwidth]{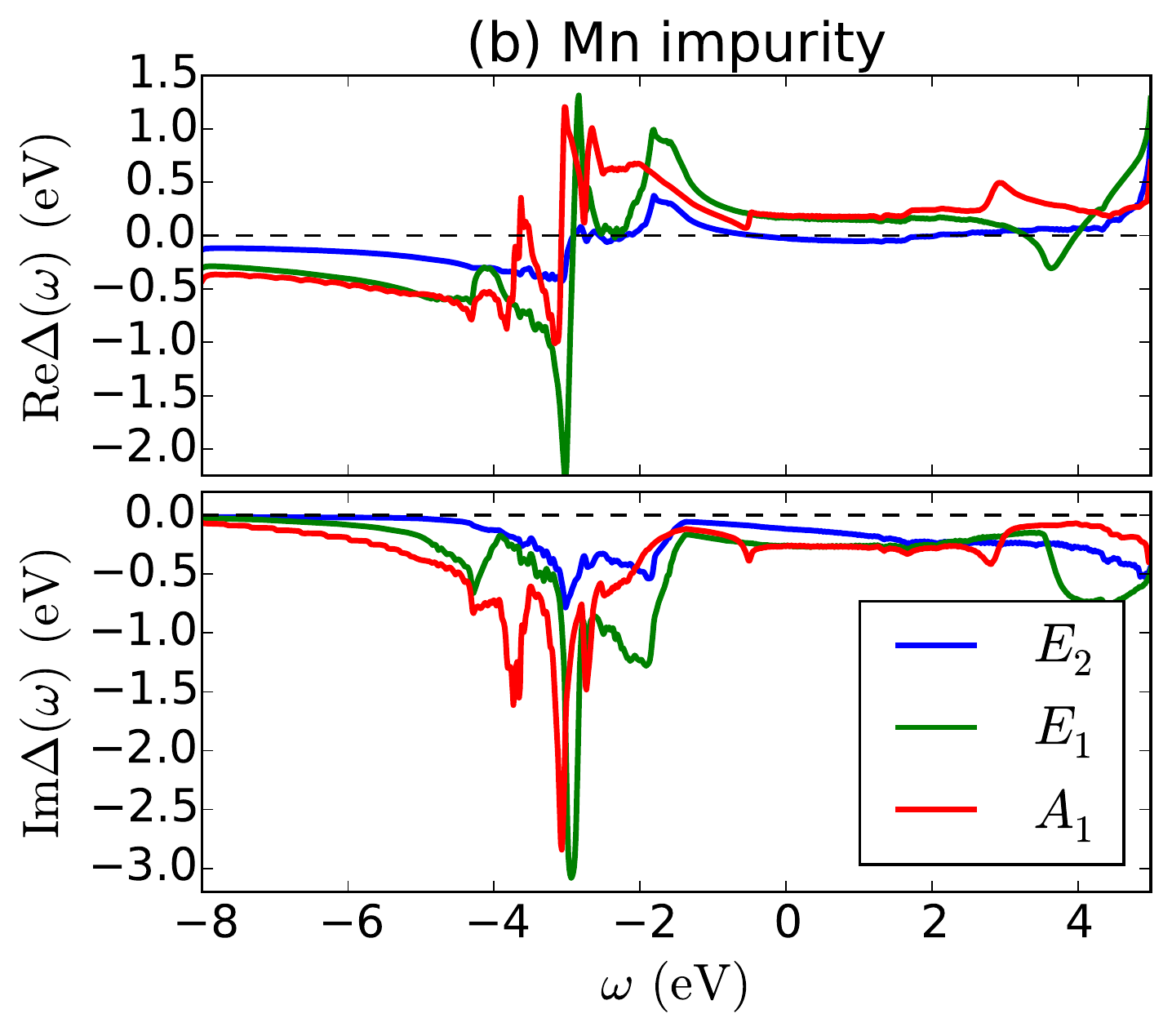}
\caption{\label{fig:hyb}(Color online) The hybridization functions, $\Delta_m(\omega)$, (real and imaginary parts) for the $d$ orbitals of Co (a) and Mn (b) adatoms on the Cu surface obtained from the KKR local DOS (c.f.~Fig.~\ref{fig:kkr_dos}).}
\end{figure}

\begin{table}[b]
\caption{\label{table:kkr_dos}
The first two rows are orbital energy level $\epsilon_m$ [Eq.~\eqref{eq:centerofmass}] and scattering rate, $\Gamma_m = -\Delta_m''(\omega=0)$, both in eV, as derived from the orbital-projected DOS of Mn and Co adatoms [Fig.~\ref{fig:kkr_dos}]. The last two rows are the corresponding Lorentzian fitting parameters for the DOS (see text).} 
\begin{center}
\begin{ruledtabular}
\begin{tabular}{ l c c c  c c c }
                   &        & Mn    &        &       & Co    &       \\
                   & $E_2$  & $E_1$ & $A_1$  & $E_2$ & $E_1$ & $A_1$ \\
                   \hline
    $\epsilon_m$   & -0.083 & -0.181& -0.054 & -0.260& -0.331& -0.234\\
    $\Gamma_m$     & 0.117  & 0.263 & 0.263  & 0.087 & 0.208 & 0.209 \\
    $\epsilon^L_m$ & -0.103 & -0.010&  0.131 & -0.264& -0.184& -0.091\\
    $\Gamma^L_m$   & 0.107  & 0.258 & 0.261  & 0.077 & 0.198 & 0.209 \\
\end{tabular}
\end{ruledtabular}
\end{center}
\end{table}

The DFT result is used to generate the input for the quantum impurity solver, the hybridization function $\Delta_m(i\omega_n)$ [Eq.~\eqref{eq:hybridization}]. Given the non-interacting Green's function $G_0(z)$ from DFT calculation ($z$ is a complex number), $\Delta_m(z)$ is obtained via Eq.~\eqref{eq:gfmatsu} (with $\Sigma_m(z) = 0$)
\begin{equation}\label{eq:hyb}
  \Delta_m(z) = z + \mu - \epsilon_m - G_m^0(\omega)^{-1}.
\end{equation}
For the non-interacting case, the zero of energy is set to the Fermi energy (or the chemical potential) of the Cu substrate. The noninteracting impurity Green function is likewise expressed in terms of the orbital-projected DOS
\begin{equation}
  G_m^0(z) = \int\!\ud\epsilon\,\frac{\rho_m^0(\epsilon)}{z + \mu - \epsilon}.
\end{equation}
The condition that $\Delta_m(z\to i\omega_n)$ decays to zero at high frequency $\omega_n$ implies that the orbital energy level is calculated via the center of mass formula
\begin{equation}\label{eq:centerofmass}
  \epsilon_m = \int\!\ud\epsilon\,\rho_m^0(\epsilon)\,\epsilon.
\end{equation}

The resulting hybridization function at real frequency is shown in Fig.~\ref{fig:hyb}.  The energy levels $\epsilon_m$ and the scattering rates, $\Gamma_m = -\Delta_m''(\omega=0)$, which quantify the hybridization strength at the Fermi energy, are given in Table~\ref{table:kkr_dos}. For comparison, we also provide the Lorentzian fitting parameters which are obtained by focusing on the main DOS peak near $E_F$ and ignoring the small spectral features near the Cu $d$ bands. 

The behavior of the hybridization functions [Fig.~\ref{fig:hyb}] follows closely that of the orbital-projected DOS [Fig.~\ref{fig:kkr_dos}]. Within the window from $\omega = -1$~eV to $+2$~eV, $\Delta_m(\omega)$ is rather smooth and featureless for all orbitals, confirming the Lorentzian-like shape of the peaks in the DOS. As mentioned previously for the orbital-projected DOS, the structures in $\Delta_m(\omega)$ around $\omega = -3$ eV come from the contribution of the Cu $d$ states on the surface. However, their contributions are unimportant. The structures above $\omega = +3$ eV  arise from free-electron-like states. Because of their high energy, they can also be safely ignored. Therefore, as seen in Appendix~\ref{app:lorentzian}, the KKR DOS as well as the hybridization function can be well approximated by the Lorentzian form around the Fermi level.

From Fig.~\ref{fig:hyb}, the hybridization functions of the Co and Mn adatoms are rather similar. The main differences can be understood in terms of the parameters given in Table~\ref{table:kkr_dos} which characterize the low-energy features of the hybridization function ($\epsilon_m$ and $\Gamma_m$). Different $\epsilon_m$ are related to different $d$ occupancies and correspond to a nearly constant shift, while the widths $\Gamma_m$ are not very different for Co and Mn. It is therefore meaningful to explore other $N_d$ values simply by varying the chemical potential while keeping $\Delta_m(\omega)$ fixed. In other words, different adatoms can be qualitatively investigated using the same hybridization functions.

\section{Correlation effect\label{sec:corr}}

Given the input hybridization function $\Delta(\omega)$ from the KKR calculation and the orbital energy $\epsilon_m$ [Table~\ref{table:kkr_dos}], we can solve the impurity model Eq.~\eqref{eq:impurity_model} exactly using the CT-HYB solver. However even when the input $\Delta(\omega)$ and $\epsilon_m$ are fixed, the impurity model still depends on the following parameters: the temperature $T$, the filling or the $d$ occupancy $N_d$ (controlled by the chemical potential $\mu$ or equivalently the double counting correction), the Hubbard value for the onsite interaction $U$ and the Hund's coupling $J$. Understanding the effects associated with these parameters is the goal of this section. As the Hund's coupling is usually unscreened and has the value of the order of $1$~eV, we keep it unchanged at $J=0.9$~eV. Therefore, we solve the impurity model [Eq.~\eqref{eq:impurity_model}] and investigate the dependence of the results on three parameters: the temperature $T$, the onsite interaction $U$ and the $d$ orbital filling $N_d$.

\subsection{Temperature effect}
For many strongly correlated metals, the main effect of decreasing the temperature is to drive the system into the Fermi liquid regime. As the Kondo temperature scale $T_K$ marks the onset of this regime, lowering the temperature and checking for the signature of Fermi liquid behavior are the proper steps to estimate the Kondo scale $T_K$. By using the CT-HYB approach, the complexity of the calculation depends on the temperature as $o(\beta^3)$ ($\beta = 1/T$) \cite{Gull11}. However, with the state-of-the-art implementation of the CT-HYB solver in the TRIQS package \cite{Seth15}, calculations become less expensive, allowing us to reach temperatures as low as $T=0.0125~\mathrm{eV}\approx 145$~K.

\begin{figure}[t]
\centering
 \includegraphics[width=\columnwidth]{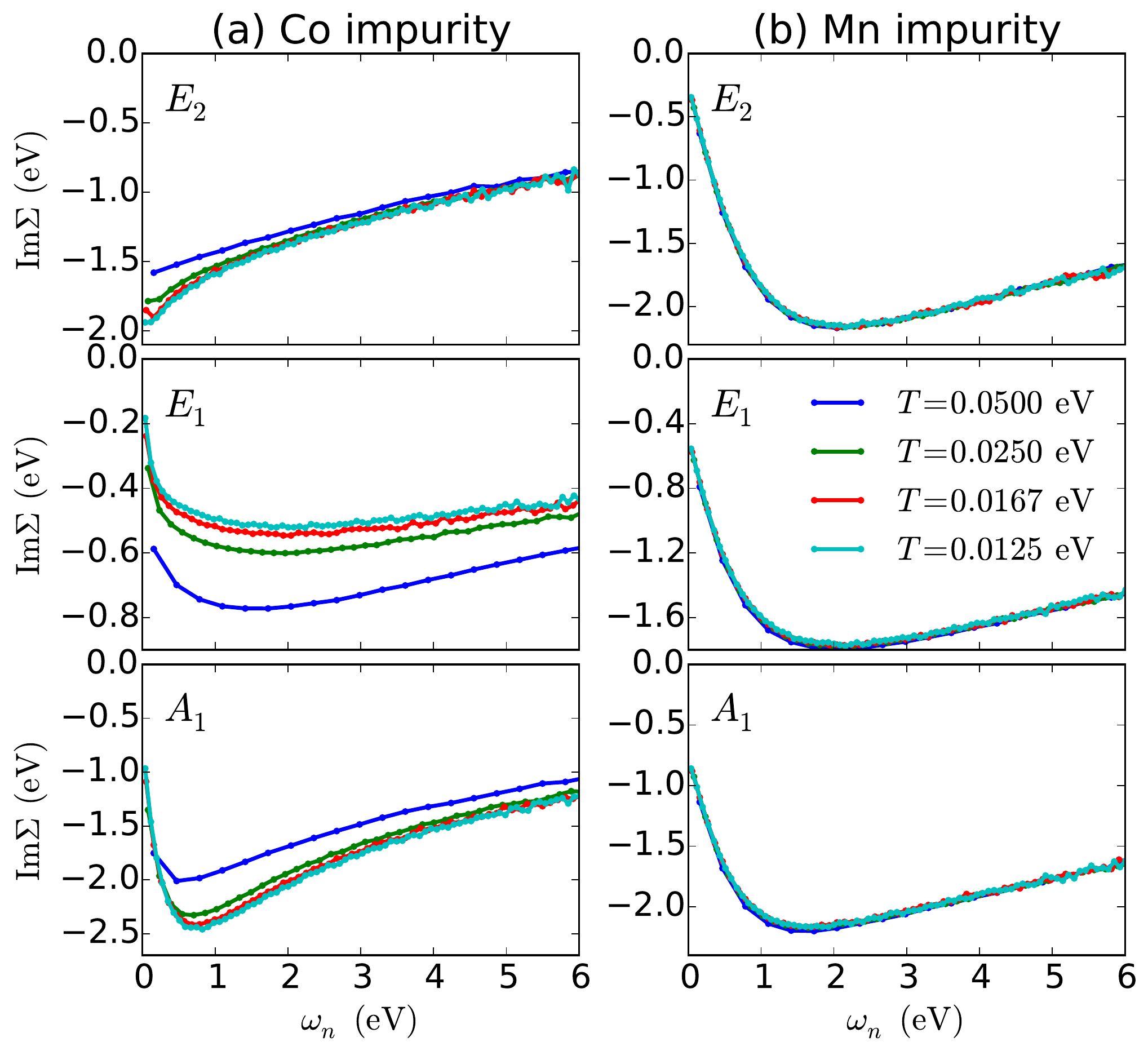}
\caption{\label{fig:temperature}(Color online) The imaginary part of the self-energy as a function of Matsubara frequency at $T=1/20,1/40, 1/60$ and $1/80$~eV for Co impurity (column a) and Mn impurity (column b). Each row corresponds to a $d$ orbital ($E_2$, $E_1$ or $A_1$). The $d$ occupancies are set at $N_d = 7.86$ for Co and $5.68$ for Mn (close to the KKR values of $N_d$). The onsite interaction is $U=4$~eV. Orbital occupancies corresponding to $E_2, E_1$ and $A_1$ at $T=1/80=0.0125$~eV are $0.53, 0.60$ and $0.56$ for Mn, $0.75$, $0.93$, and $0.57$ for Co.}
\end{figure}

Figure~\ref{fig:temperature} shows the self-energies for Co and Mn adatoms for a wide range of temperatures from $0.0125$ to $0.05$~eV, with $N_d$ fixed at $7.86$ and $5.68$ for Co and Mn adatoms, respectively, which are not far from the KKR occupancies. The self-energies behave differently for the two adatoms as the temperature decreases. For Mn, the self-energies for different temperatures are almost on top of each other, while for Co, the self-energy for each orbital gradually converges as the temperature decreases to the value $T = 0.0125$~eV. This difference is understood from the orbital polarization when lowering the temperature. For Mn, the orbital occupancies (see the caption of Fig.~\ref{fig:temperature}) are almost unchanged for the different temperatures in use. As a result, the self-energies of the Mn adatom at different temperatures are nearly unchanged. On the other hand, the Co adatom exhibits clear changes in orbital polarization. As the temperature decreases, the $E_2$ and $A_1$ occupancies decrease while that of $E_1$ increases. The occupancies nearly converge when $T\to0.0125$~eV (values are given in the caption of Fig.~\ref{fig:temperature}). This explains the changes in the corresponding self-energies of the Co adatom. The overall difference comes from the fact that Mn is close to half-filling. Thus the Hund's coupling effect is strong in keeping the high spin state and reducing orbital ordering. It is therefore less likely to observe orbital ordering unless the value of $J$ decreases.

The difference between the self-energies of the three orbitals ($E_2$, $E_1$ and $A_1$) also deserves attention. First, the frequency regime in Fig.~\ref{fig:temperature} can be divided into two parts separated by the frequency at the self-energy pole ($\sim 1$~eV for Co and $\sim 2$~eV for Mn adatom). At large frequencies, the correlation strength is controlled by the Hubbard $U$ and depends on how close the orbital occupancy is to $0.5$ (half-filling) \cite{deMedici2014}. According to the orbital occupancy values of Co provided in the caption of Fig.~\ref{fig:temperature}, $A_1$ is the most correlated orbital due to its occupancy closest to $0.5$, followed by $E_2$ and $E_1$; for Mn, $E_1$ is the least correlated orbital because it has the largest occupancy. The high frequency correlation strengths of $E_1$ and $A_1$ are similar as their occupancies are not very different.

The low frequency part of the self-energy is more important as it is related to the low-energy physics near the Fermi level and determines the Kondo temperature scale $T_K$. At low frequency, the Hund's coupling $J$ may play an important role. Thus the orbital which is more localized near the Fermi level is more correlated \cite{Mravlje11}. Consequently, for Co, the $E_2$ orbital which has the smallest hybridization (see Table~\ref{table:kkr_dos}) is the most correlated, followed by the $A_1$ and $E_1$ orbitals. However, the interpretation may not be universal, for example, when considering the Mn case as presented in Fig.~\ref{fig:temperature}. We believe that other factors, such as the proximity to half-filling, the orbital ordering, or the too high temperature may complicate the situation. Understanding the low-frequency self-energy is still an open problem for future investigation.

With a wider range of temperatures under investigation, we achieve better estimates of the Kondo temperature than previous studies \cite{Surer12}. Using Eq.~\eqref{eq:kondoT}, as in Ref.~\onlinecite{Surer12}, we obtain the Kondo temperatures for all orbitals. The largest $T_K$ is associated with orbital $E_1$, which has occupancy away from half-filling and a large scattering rate (see Table.~\ref{table:kkr_dos}). For Co, the $T_K$ value of the $E_1$ orbital converges at $0.023$~eV, while for Mn, $ T_K $ decreases slowly as the temperature is lowered and reaches $0.009$~eV at $T=0.0125$~eV. However, the fact that the experimental value of the Kondo temperature is approximately $\sim 0.005$~eV for Co~\cite{Knorr2002,Manoharan00} and a well-formed local moment is observed for Mn~\cite{Gardonio13,Nevidomskyy2009} suggests that the total $d$ occupancy around the DFT value is not a good choice. The adatom $d$ electrons may therefore be more localized than predicted by the DFT calculations. We will analyze this issue later in the paper.
  
To explore temperatures below those accessible within CT-QMC, we have also used ED as multi-orbital impurity solver, with $T$ in the range $0.0025 \to 0.025$~eV. As discussed in Appendix B, we find a characteristic trend, namely that the initial slope of the self-energy of all orbitals increases when $T$ decreases. Thus, the quasi-particle weights $Z_m$ and Kondo temperatures $T_{K,m}$ decrease. These results suggest that, at the temperatures used in the calculations, the system has not yet reached the true Fermi-liquid regime. We point out, however, that although this temperature trend is physically reasonable, the results must be used with caution because of finite-size limitations inherent in ED. As illustrated in Appendix B, using only two bath levels per impurity orbital is no longer adequate to accurately fit the non-interacting Green's function at very low $T$, even if the DOS components are assumed to have Lorentzian shape. Thus, the low-energy behavior of the self-energy at these low values of $T$ does not permit a reliable determination of the quasiparticle weights $Z_m$ and of the corresponding Kondo temperatures. Using CT-QMC or ED is therefore presently not possible to reach the actual temperatures used in the experimental STM studies.

\begin{figure}[t]
\centering
 \textbf{Co adatom} \\
 \includegraphics[width=\columnwidth]{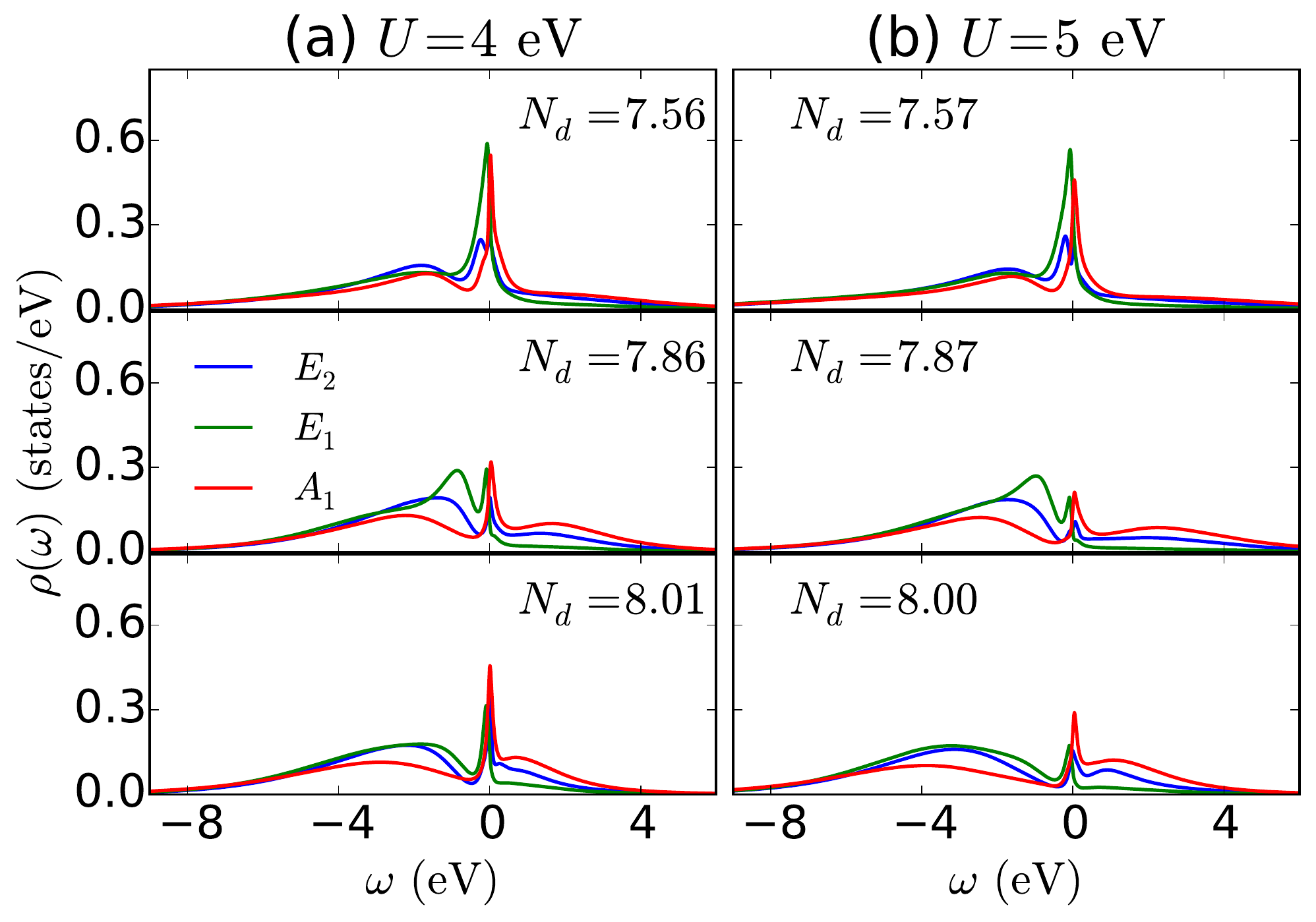} \\
 \textbf{Mn adatom} \\ 
 \includegraphics[width=\columnwidth]{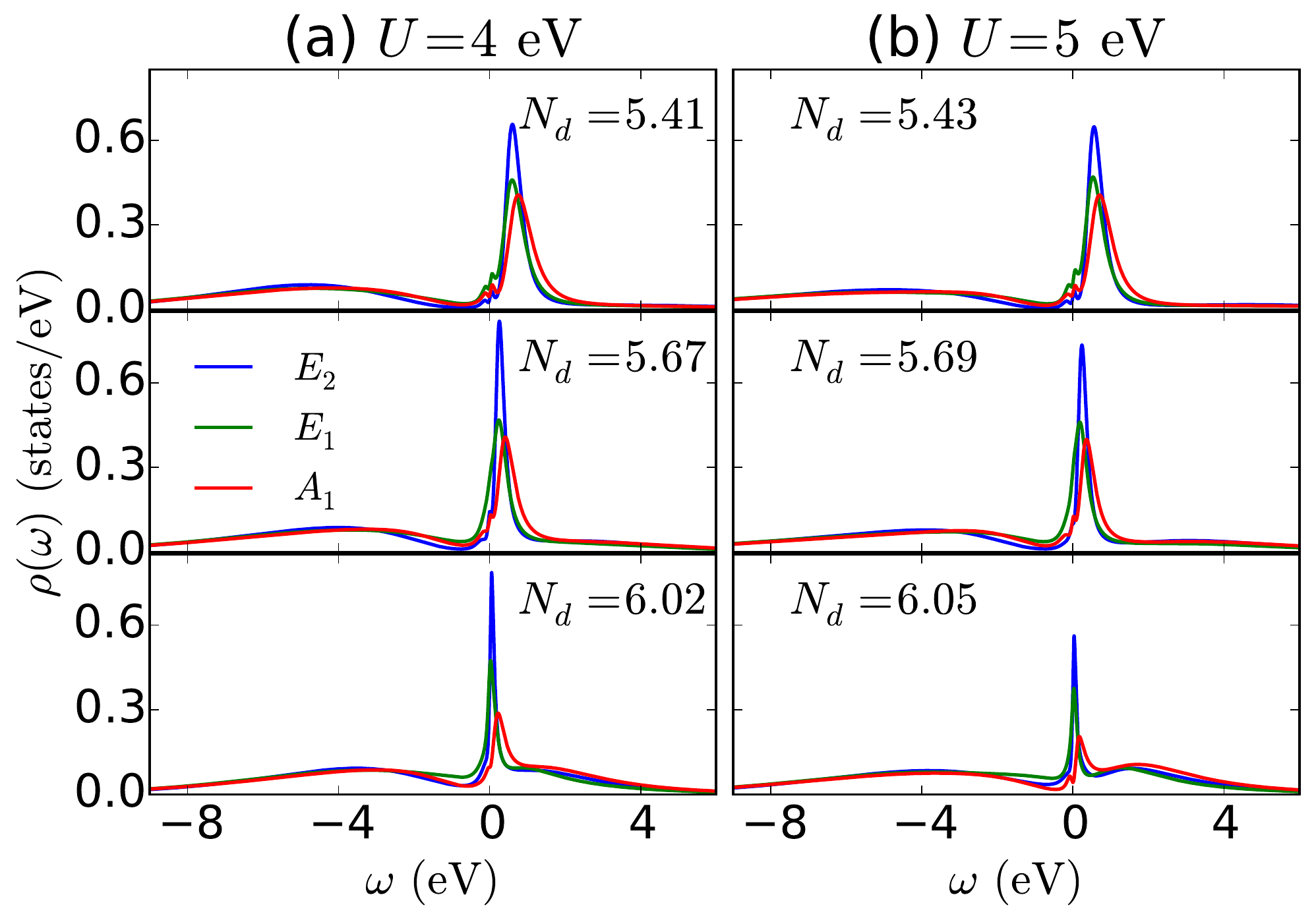}
\caption{\label{fig:CoMnU45} (Color online) The orbital-projected DOS $\rho_{m}(\omega)$ for Co (top) and Mn (bottom) adatoms for $U=4$ (column a) and 5 eV (column b) at three different total $d$ occupancies $N_d$. The temperature is $T=0.025$~eV.}
\end{figure}

\subsection{Interaction strength ($U$ and $J$)}

The second parameter, the interaction strength, drives the electronic correlations; the investigation of its impact is therefore crucial. Here we consider two values, $U=4$ and $5$~eV, which are close to those used in the literature~\cite{Jacob09,Surer12,Gardonio13}. As mentioned previously, the Hund's coupling $J$ is kept fixed. We focus on the question of how the Kondo temperature and the spectra are affected by the increase of $ U $ within the physical range.

Figure~\ref{fig:CoMnU45} shows the many-body DOS $\rho_m(\omega)$ for $U=4$ and $5$~eV produced using MaxEnt for the analytic continuation. The peak at the Fermi level (the Kondo peak) represents the magnitude of the Kondo temperature $T_K$; in other words, it tells us how correlated the impurity system is. In both panels of Fig.~\ref{fig:CoMnU45}, the difference of spectra at different $U$ is very small when the total occupancy is away from integer filling ($N_d=7.56$ for Co or $5.41, 5.67$ for Mn). At occupancy values closer to integer filling ($N_d=7.86, 8.01$ for Co or $N_d=6.02$ for Mn), the magnitude of the Kondo peak is seen to decrease as the onsite interaction $U$ increases. Consequently, the renormalization factor $Z$ and the Kondo temperature $T_K$ (not shown) decrease in the same way.

The physics associated with this behavior is understood by comparing the adatoms to strongly correlated bulk materials, where a metal-insulator transition is observed as a function of doping (filling-control metal-insulator transition \cite{Imada98}). At integer filling, hopping of electrons between sites costs an energy related to the onsite interaction, while in doped systems, electron hopping only costs an energy of the hopping amplitude, which is typically much smaller. Thus an energy gap $\sim U+(M-1)J$ for half-filling or $\sim U-3J$ for non-half-filling ($M$ is the number of electrons per site) is formed for bulk materials at large $U$ and at integer filling \cite{Georges13}. Similarly, for the impurity model at integer filling, the interaction $U$ forms an energy barrier at the impurity that forbids the background electrons to move in. At integer filling, the large $U$ suppresses the charge fluctuation and allows to map the system into the Kondo model with the Kondo antiferromagnetic coupling $J_s\sim \sum|V_{\mathbf{p}i\sigma}|^2/U$. As $T_K \sim \exp(-1 / J_s)$, increasing $U$ implies that the Kondo temperature scale decreases. On the other hand, for non-integer filling, the charge fluctuation is strong and reduces the correlation effect. $T_K$ and $Z$ are then less affected by the change in $U$. This explains why there is almost no change in the spectral functions for non-integer cases at different $U$ values, while there is a decrease in the Kondo peak for cases close to integer filling.

We also point out that, for Co (upper panel of Fig.~\ref{fig:CoMnU45}), the lower and upper Hubbard satellites are close to the Fermi level with centers at $\sim -2$ and $1$~eV. They are most clearly observed at occupancies close to integer filling ($N_d \sim 7.86$ or $8$). The small energy gap for charge excitation $\sim U-3J$ \cite{Georges13} explains why the Hubbard satellites are close to the Fermi level. A small increase of the Hubbard band separation for Co is found when $U$ increases from $4$ to $5$~eV. For Mn (lower panel of Fig.~\ref{fig:CoMnU45}), at $N_d \sim 5.4$, the lower Hubbard band is centered farther below the Fermi level at $\lesssim-4$~eV while the upper Hubbard band is the large peak near the Fermi level when the system is close to half-filling $N_d=5$. This accounts for larger charge energy gap. Increasing the filling to $N_d\sim 6$ transforms the spectral portion of the upper Hubbard band into the Kondo peak, leaving a smaller upper Hubbard band centered at $\sim +2$~eV.

\subsection{Filling effect}

The influence of temperature and interaction effect on the impurity system depends strongly on whether the $d$ occupancy $N_d$ is integer or not, and, in particular, on whether it is close to or far from half-filling. As noted above, the ill-defined double counting correction makes the precise value of occupancy $N_d$ not well determined. Therefore, this subsection is devoted to the detailed analysis of the role of the occupancy $N_d$. For this purpose, we fix other parameters ($U=4$~eV, $J=0.9$~eV and $T=0.05$~eV) and vary $N_d$ in a wide range to study its effect. 

\begin{figure}[t]
\centering
 \includegraphics[width=0.49\columnwidth]{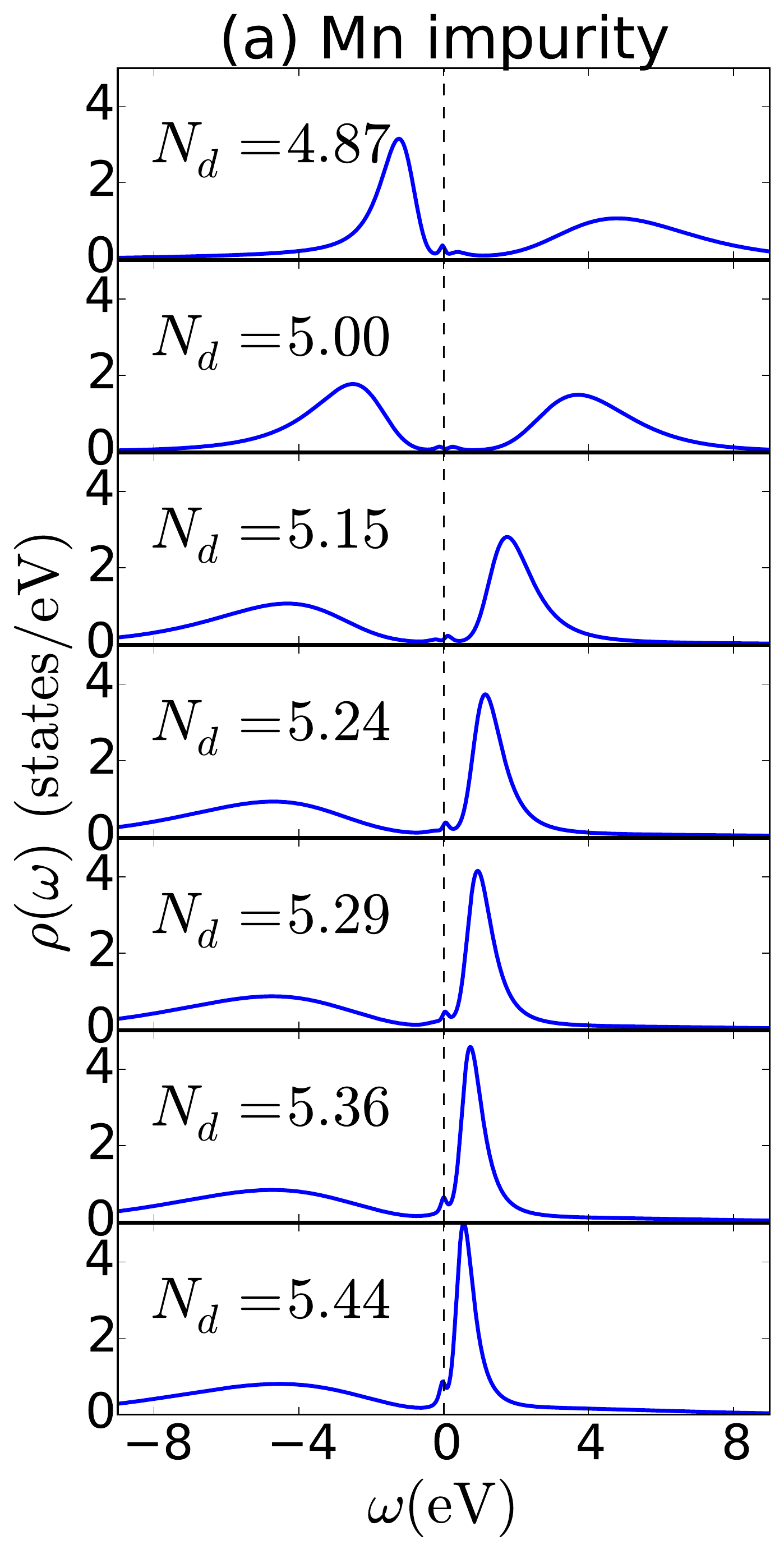}
 \includegraphics[width=0.49\columnwidth]{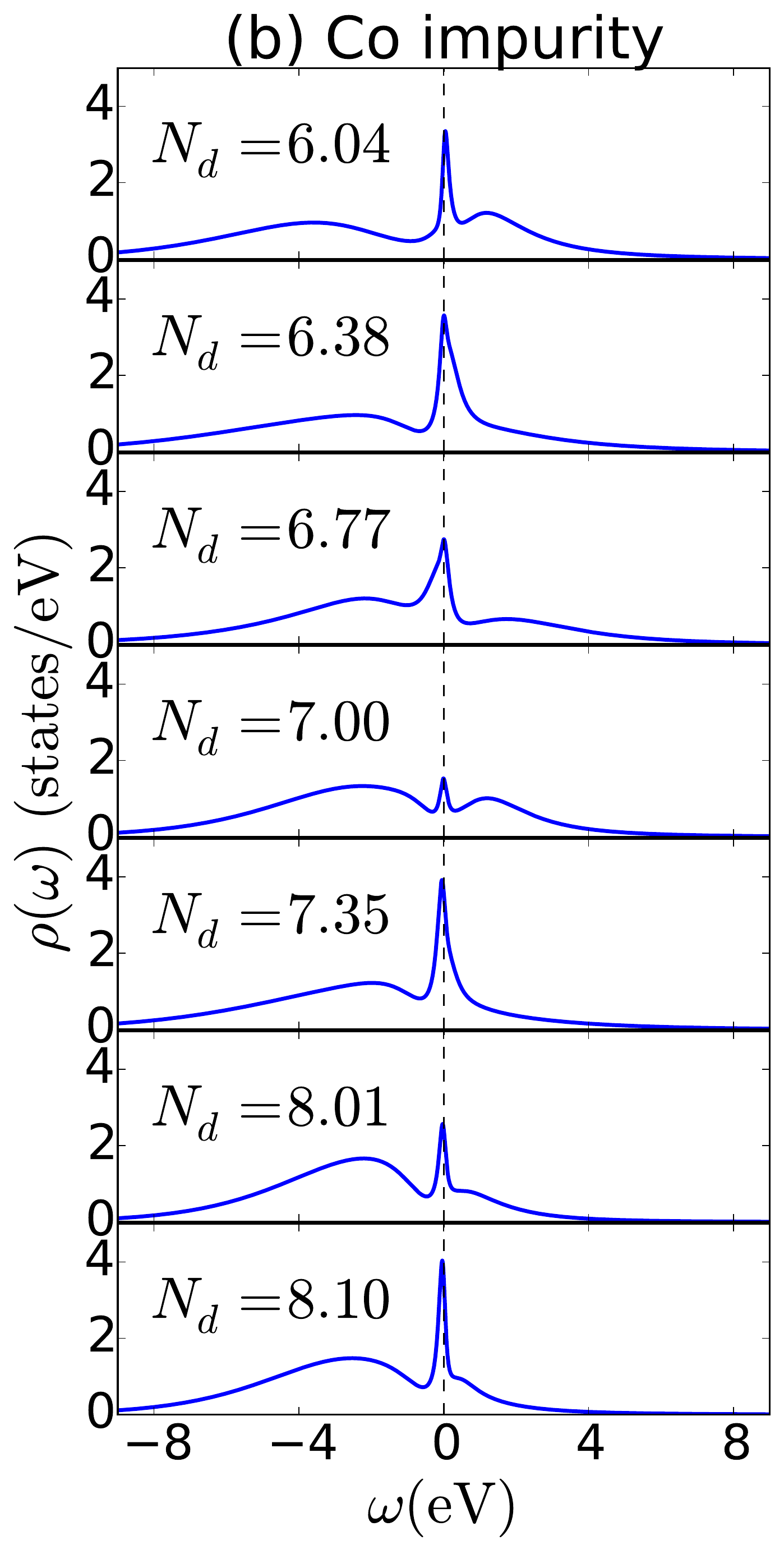}
\caption{\label{fig:dos_evolution}
The evolution of spectral functions $\rho_{m}(\omega)$ with increasing filling $N_d$ for Mn (a) and Co (b). The onsite interaction parameters are $U=4$~eV, $J=0.9$~eV and the temperature is $T=0.05$~eV.}
\end{figure}

\begin{figure*}[t]
\centering
 \begin{tabular}{c}
 \includegraphics[width=\columnwidth]{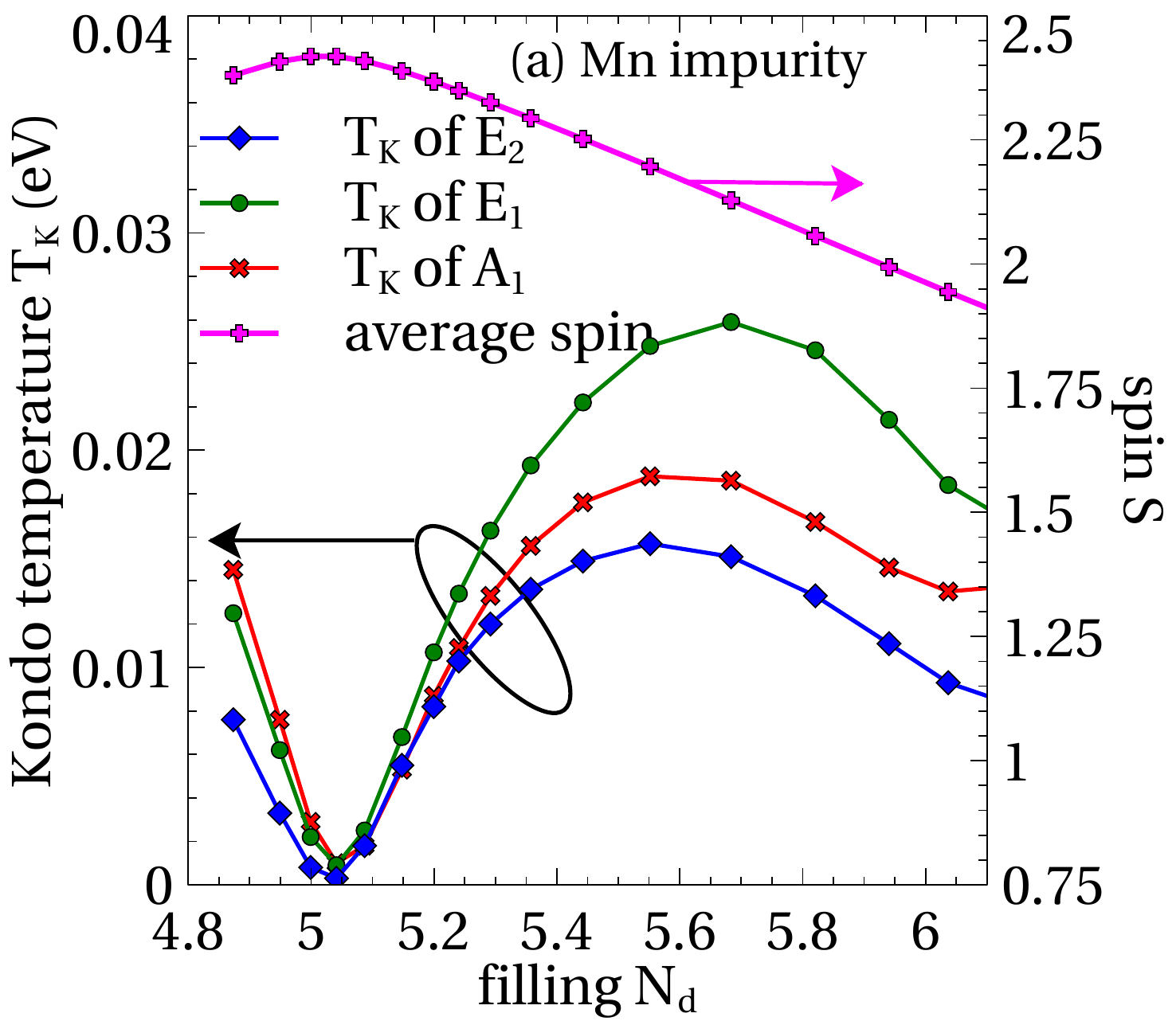}
 \includegraphics[width=\columnwidth]{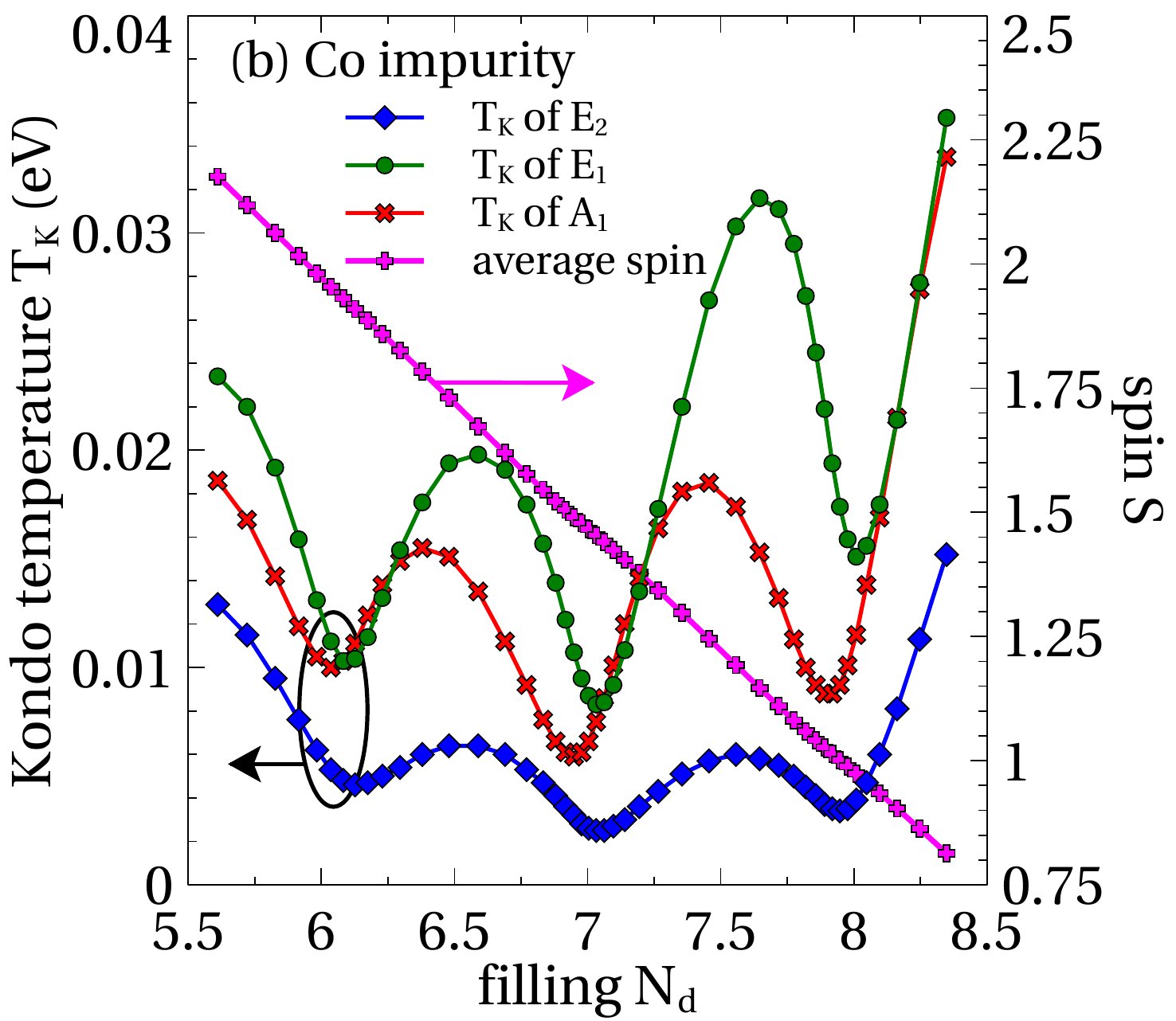}
 \end{tabular}
\caption{\label{fig:TkSplot}(Color online)  The dependence of the Kondo temperature $T_K$ and the impurity spin $S$ on the filling $N_d$. The Hubbard value is $U=4$~eV and the temperature is $T=0.05$~eV.}
\end{figure*}

Figure~\ref{fig:dos_evolution} shows the evolution of the spectra as the $d$ occupancy increases. For the Mn adatom, $N_d$ is varied from below half-filling to about $6$. The corresponding spectra are shown in Fig.~\ref{fig:dos_evolution}(a). Around half-filling, the spectral weight at the Fermi level is depleted. As $N_d$ increases to $6$, some weight from the upper Hubbard band is transferred to the DOS near the Fermi level and the Kondo peak is seen to develop gradually. The evolution of the spectra for the Co adatom is presented in Fig.~\ref{fig:dos_evolution}(b). In this case, the changes are less pronounced than those for Mn. As the filling is far from half-filling, a Kondo peak is obtained for all occupancies. However, the magnitude of this peak decreases as $N_d$ becomes integer and increases again when it is away from integer values.

The evolution of the aforementioned spectra is reflected in the $T_K$ versus $N_d$ plot in Figure~\ref{fig:TkSplot}. Note that $T_K$ is calculated at temperature $T=0.05$~eV, which is not low enough for a quantitative determination of $T_K$. However, it provides a qualitative understanding of how $T_K$ depends on the $d$ occupancy. Also, at $N_d$ where the two cases overlap, Mn has a slightly larger $T_K$ than Co, which is mainly due to the larger hybridization $\mathrm{Im}\Delta$ at the Fermi level (Table~\ref{table:kkr_dos}). Nevertheless, the tendencies of $T_K$ are similar for the two impurities. This again confirms that the same hybridization function can be used for both adatoms and that the $d$ occupancy of the impurity is an important quantity. Consequently, Fig.~\ref{fig:TkSplot} can be considered as a universal plot of the dependence of $T_K$ on $N_d$ in a wide range from $N_d=5$ to $8$.

The minima of $T_K$ near $N_d=5,~6,~7,$ and $8$ are clear evidence for the fact that integer filling suppresses the Kondo scale. At integer filling, charge fluctuation largely decreases due to strong correlation and the contribution to the Kondo peak is mainly from spin fluctuations. Hence, minima in $T_K$ occur at integer filling. Away from integer filling, correlation effects become weaker, so that electrons can hop between the impurity and the host material without much cost of energy due to the interaction. Thus, charge fluctuations contribute more to the formation of the Kondo peak and, together with the spin and orbital fluctuations, they enhance $T_K$. We note that, because of the high temperature used in the CT-QMC impurity solver, the minima of $T_K$ slightly deviate from integer fillings. Our calculations using the ED impurity solver (not shown) indicate that at lower temperatures the minima are in better agreement with integer occupancies.

Furthermore, at integer filling, the system can be well approximated by the traditional Kondo model \cite{book:Hewson97} where there is only coupling $J_s$ between the impurity spin and the spin of itinerant electrons. The impurity spin is screened more easily if its value $S$ is small (see e.g. Ref.~\onlinecite{Georges13} and references therein). Thus in Fig.~\ref{fig:TkSplot}, $T_K$ reaches the minimum at half-filling where $S=5/2$ is maximal. $T_K$ becomes larger at $N_d=7$ and $8$ because of the smaller impurity spin. We notice that $N_d=6$ is a special case with larger $T_K$ than at $N_d=7$, which may be due to weaker Hund's coupling at this $N_d$. It is an interesting open problem to fully understand this special case.

By studying a wide range of $d$ occupancy and comparing with experimental data, it might be possible to define a value of $N_d$ that matches the experiment. As mentioned previously, the DFT value of $N_d$ may not be a good choice for the impurity system. Consider first the Mn adatom. Experiments of Mn in bulk alloys \cite{Nevidomskyy2009} and on Ag(100) surface \cite{Gardonio13} suggest it have a well-formed local moment. Thus the Kondo peak at the Fermi level is suppressed \cite{Gardonio13, Nevidomskyy2009}. According to Fig.~\ref{fig:dos_evolution}(a), $N_d \sim 5$ appears appropriate. For the Co adatom, to match the experimental Kondo temperature $T_K \sim 0.005$~eV \cite{Knorr2002,Manoharan00}, $N_d$, according to Fig.~\ref{fig:TkSplot}(b), should be closer to $7$. From our calculations at lower temperatures (not shown), $N_d$ is larger than $7$ but should be smaller than the DFT value $7.85$. Therefore, as a result of interaction, the $d$ orbitals of the impurity become less hybridized, so that its total occupancy is closer to the formal valence. This finding is similar to  DMFT studies of bulk transition metal oxides \cite{Haule2014,Dang14}, which show that the double counting correction should be chosen such that the $d$ occupancy of the correlated transition metal atom is close to its formal valence. However, we note that for the case of the Co adatom, the main physics does not change qualitatively within the range of $N_d$ between $7$ and $8$. We therefore choose $N_d\approx7.85$ for the calculations discussed in the next section, Sec.~\ref{sec:stm}.

\section{Theoretical STM spectra\label{sec:stm}}

STM experiments do not probe directly the local DOS of the adatom but the electronic states in the vacuum region surrounding it. The results of the previous sections help us to obtain the modification of this electronic structure due to correlations within the adatom 3$d$ shell. The connection to the STM spectra is established within the Tersoff-Hamann model as indicated in Eq.~\eqref{eq:rhos}.

The new ingredient in the present work is the $d$-electron self-energy. We select two cases to illustrate the analysis of the STM spectra: Mn with $N_d \approx 5.04$ and Co with $N_d \approx 7.85$, where $U = 4$~eV and $T = 0.025$~eV. These $N_d$ values are close to those used in other works \cite{Surer12,Gardonio13,Khajetoorians2015}. As can be seen in Figs.~\ref{fig:CoMnU45} and \ref{fig:dos_evolution}, the local DOS related to these values represent well the local moment limit for the Mn adatom and the Kondo physics for the Co adatom.

\begin{figure}[t!]
\centering
\includegraphics[width=\columnwidth]{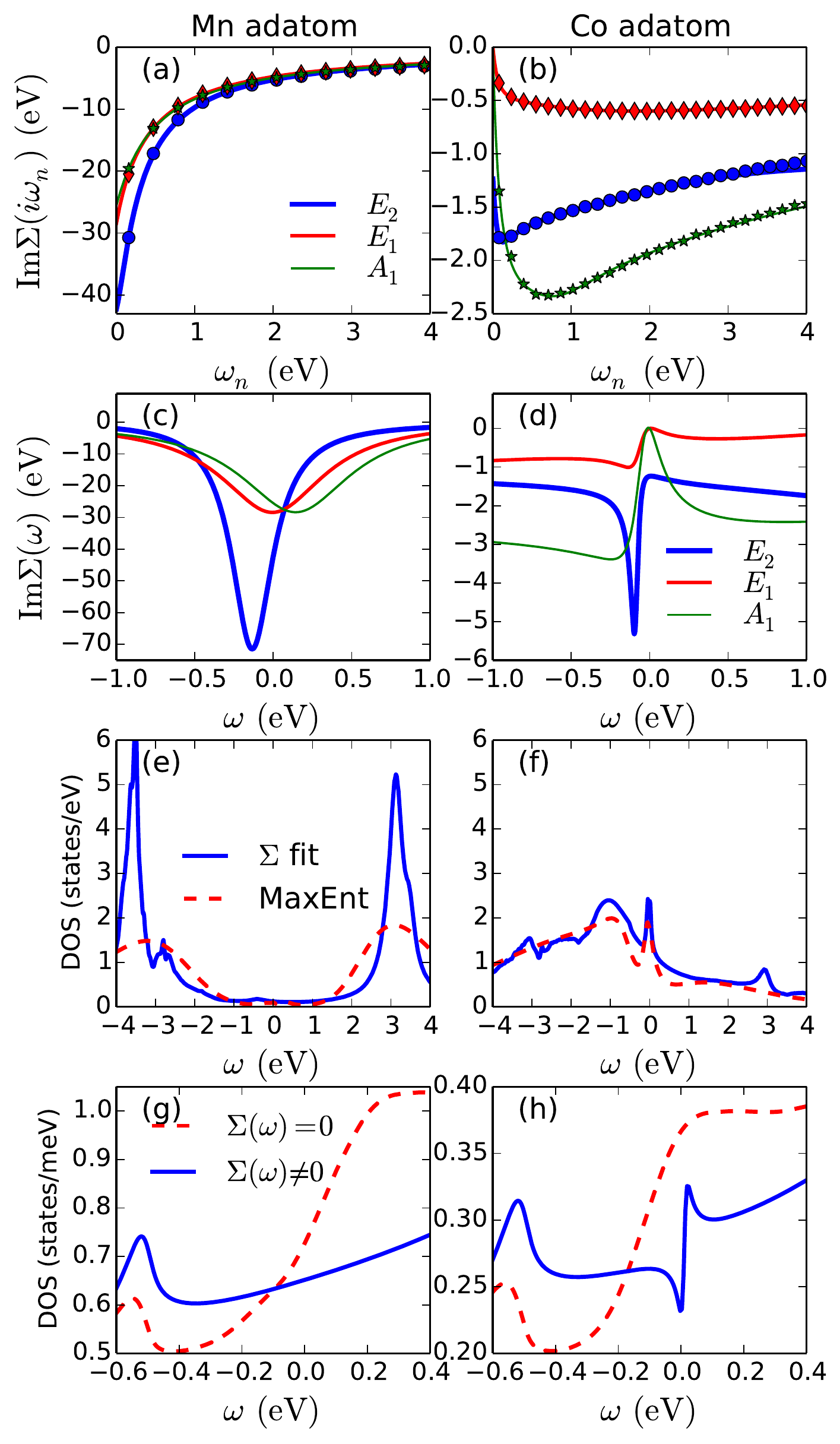}
\caption{\label{fig:sefit}(Color online) First line: comparison of the rational fit function [Eq.~\eqref{eq:ratfit} with $N=3$] (solid lines) to the CT-QMC data for $\IM\,\Sigma(\iu\omega_n)$ (symbols) for Mn (a) and Co (b). Second line: analytical continuation of the CT-QMC $\Sigma(\iu\omega_n)$ to real frequency $\mathrm{Im}\Sigma(\omega+i0)$ using the rational fit function [Eq.~\eqref{eq:ratfit}] for Mn (c) and Co (d). Third line: local $d$ DOS for the adatoms, via MaxEnt and by solving the KKR Dyson equation with the results from (c) and (d) for Mn (e) and Co (f). Fourth line: STM spectra in vacuum $4.2$~\AA~vertically above the adatom [Eq.~\eqref{eq:rhos}] for Mn (g) and Co (h). The impact of the self-energy is highlighted by comparing the DOS before and after the self-energy is taken into account. The following parameters are used: $U = 4$~eV, $J=0.9$~eV, $T = 0.025$~eV at $N_d = 5.04$ for Mn and $N_d = 7.85$ for Co.}
\end{figure}

Figure~\ref{fig:sefit} shows the step-by-step results towards STM spectra for the two cases. The CT-QMC self-energy and the fitting function on the Matsubara axis are shown in Figs.~\ref{fig:sefit}(a) and (b), with the analytical continuation to real frequency shown in panels (c) and (d). Our interest is in the low-frequency region around the Fermi level. As the Mn adatom is near half-filling, the self-energy acts to create a gap in the DOS. $\RE\,\Sigma$ pushes states away from the Fermi energy, and $\IM\,\Sigma$ is very large, implying a very short lifetime near $E_F$. On the other hand, the qualitative behavior of the Co self-energy is orbital-dependent. $\IM\,\Sigma$ for the $E_1$ and $A_1$ orbitals seems to approach zero for $\omega = 0$, contrary to the $E_2$ case. This means that, for $T = 0.025$~eV, the former already display the anticipated low-frequency Fermi liquid behavior, while the latter would require a much lower simulation temperature for a crossover into that regime. The local DOS peak at $E_F$ is thus generated from the contributions of the $E_1$ and $A_1$ self-energies, where $\RE\,\Sigma \propto (1-Z^{-1})\omega$ leads to narrowing of spectral features and $\IM\,\Sigma \propto \omega^2$ to small broadening. A similar orbital-dependent behavior can be observed in Ref.~\onlinecite{Surer12} and has been recently discussed in Ref.~\onlinecite{Khajetoorians2015}.

\begin{figure}[t]
\centering
\includegraphics[width=\columnwidth]{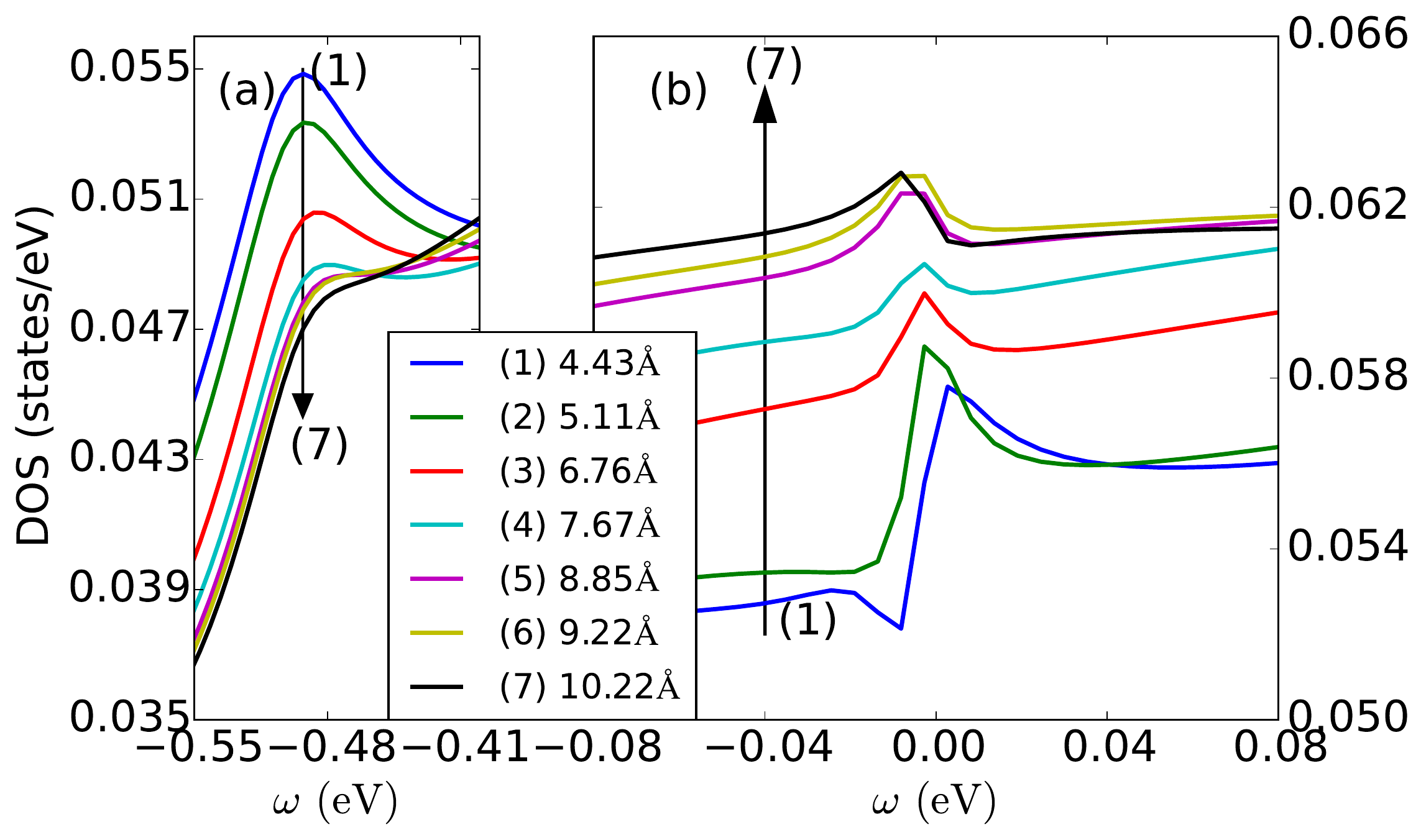}
\caption{\label{fig:surfdos}(Color online) Surface DOS averaged at different positions away from the Co adatom, from Eq.~\eqref{eq:rhos}. (a) Bound state near the onset energy for the Shockley surface state. (b) Fano-like behavior near $\EF$ arising from the strong correlations at the adatom. The distance away from the adatom is increased along the direction of the arrow.}
\end{figure}

\begin{figure}[t]
\centering
  \includegraphics[width=0.5\columnwidth]{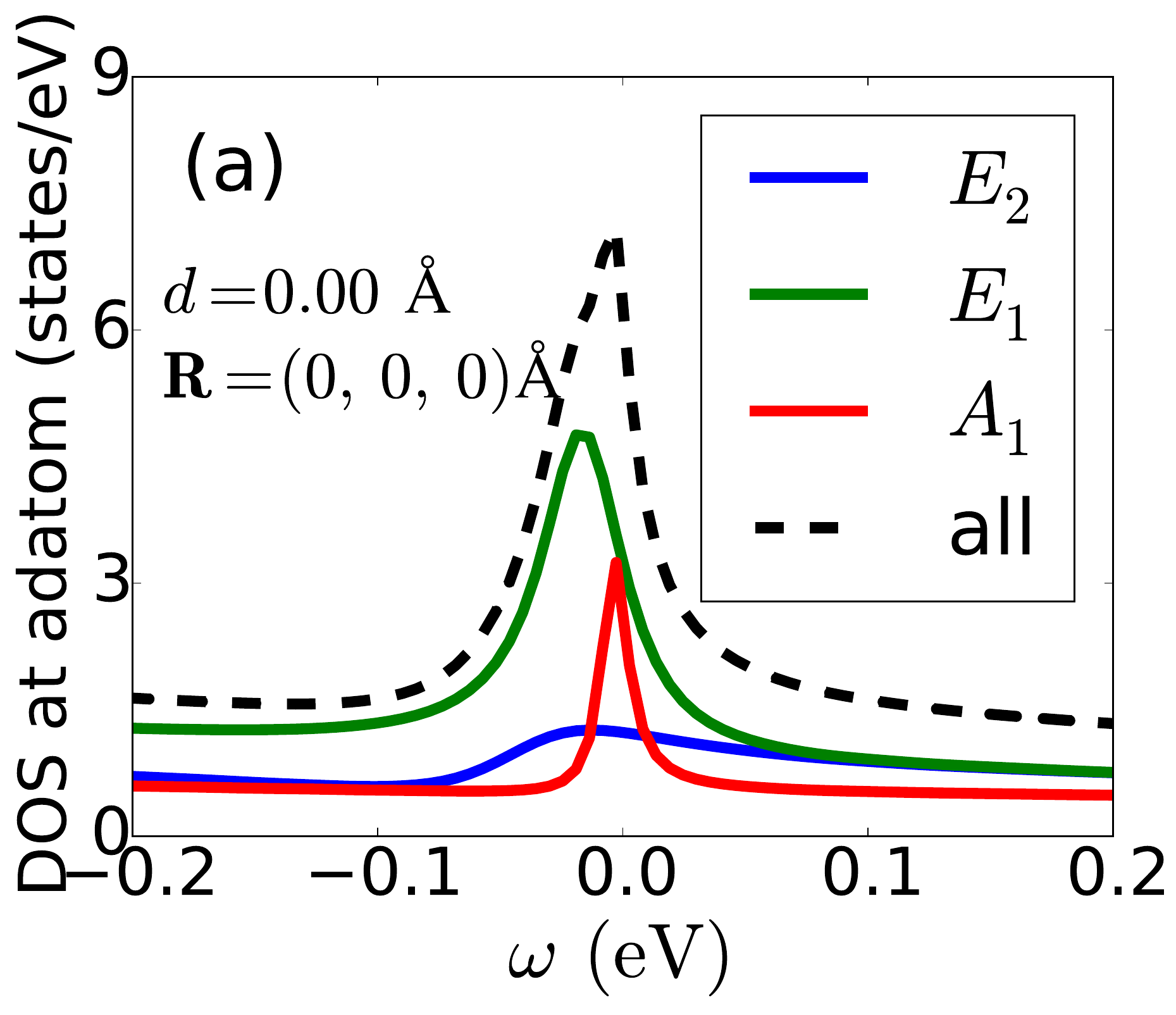} \\ 
  \includegraphics[width=\columnwidth]{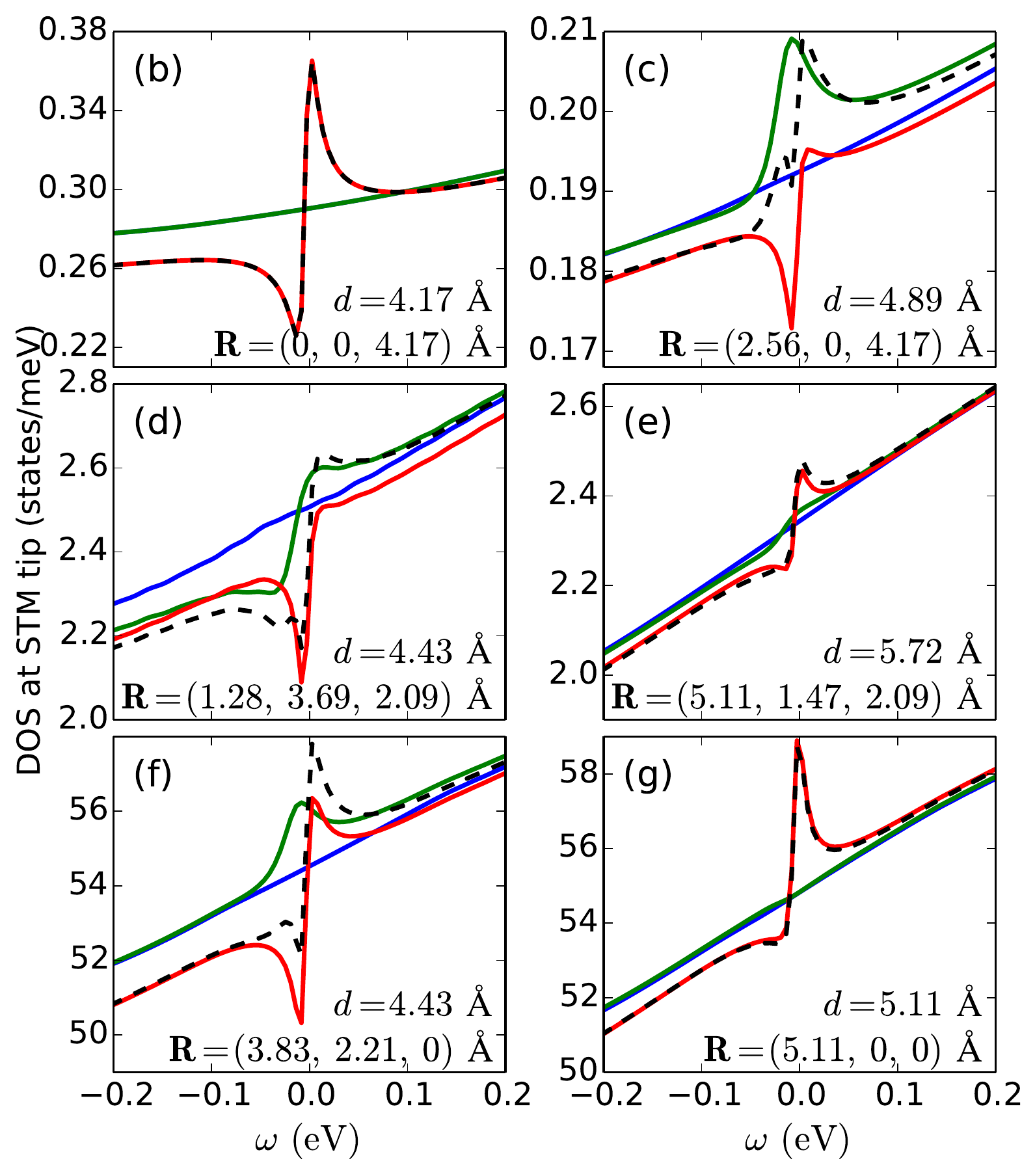}
\caption{\label{fig:surfdos2} (Color online) Contributions from different correlated orbitals ($E_2 = \{x^2-y^2, xy\}$, $E_1 = \{xz,yz\}$ and $A_1 = \{3z^2-r^2\}$) to the Fano line shapes in the surface DOS of Co. The adatom is taken as origin of the tip coordinates ($\mathbf{R}_{\mathrm{Co}}=(0,0,0)$); the Cu(111) surface plane corresponds to $z = -1.88$~\AA. The different curves show the effect of including the self-energy of only one orbital on the surface DOS around the adatom. The black dashed curve is the spectrum in the case where the self-energies of all orbitals are kept. The legend in panel (a) is applied to all other panels.}
\end{figure}

Figures~\ref{fig:sefit}(e) and \ref{fig:sefit}(f) display the local DOS at the adatoms, summed over the five orbitals. The MaxEnt analytical continuation of the QMC data is compared with the DOS obtained from updating the KKR Green's function with the rational fit to the self-energy. The overall features are in good agreement, but the MaxEnt data shows less structure. The Hubbard peaks for Mn are enhanced with the rational fit to the self-energy, so is the quasiparticle peak at $\EF$ for Co. The latter enhancement arises from assuming $\IM\,\Sigma(0) = 0$ for $E_1$ and $A_1$ when constructing the rational approximation. The former may arise from an incorrect high-frequency behavior of the rational approximation, as it was designed for low frequency. Figures~\ref{fig:sefit}(g) and \ref{fig:sefit}(h) show the DOS $4.2$~\AA~vertically above each adatom, as seen via the tip for two cases: with and without the inclusion of the self-energy. Within the LDA, both Mn and Co exhibit broad peaks near $\EF$. When the strong correlation is treated dynamically, the vacuum DOS above Mn becomes featureless near $\EF$, while for Co the signature of the quasiparticle peak takes the characteristic form of a Fano line shape. We also note that there is a stable feature in the surface DOS near $0.5$~eV below $\EF$, visible in Figs.~\ref{fig:sefit}(g) and \ref{fig:sefit}(h), which is present with and without the self-energy. This feature was explained in Refs.~\onlinecite{Limot2005,Lounis2006} as a bound state split off below the bottom of the Cu(111) surface state band, induced by the presence of the adatom.

To make more specific contact with experimental conditions, we now consider a variety of tip positions. First, we perform a lateral scan, keeping the same height above the surface as the adatom (see Fig.~\ref{fig:impurity_cartoon} for the investigated positions). The results are shown in Fig.~\ref{fig:surfdos}. The peak at $\omega=-0.5$~eV is a useful reference; its intensity decays as a function of the distance from the adatom in a monotonic way. In contrast, the Fano line shape near $\EF$ has a more interesting distance dependence, switching progressively from step-like to peak-like to a reversed step.

Next, we disentangle the contributions of different correlated orbitals to the theoretical STM spectra. As different orbitals have different symmetries with respect to the surface, their hybridization functions and self-energies give rise to different weights in the local DOS at the tip. We select six tip positions for this analysis: three close to the adatom, and three further out. 

Figure~\ref{fig:surfdos2} shows the computed STM spectra using the self energy obtained for the Co adatom with $U=4$~eV, $J=0.9$~eV and $T=0.025$~eV for several STM tip positions. In each case, the STM spectrum per orbital is calculated such that only the self-energy corresponding to that orbital is kept while those of other orbitals are suppressed. The spectrum in which all orbital self-energies are maintained (black dashed curves in Fig.~\ref{fig:surfdos2}) is also plotted for reference. The top row [panel (a)] shows the contribution from each orbital to the local DOS of the adatom (the tip is at the position of the adatom). Thus it cannot be probed directly by STM. The spectra show that the $E_1$ orbital is the most prominent, followed by $A_1$, while $E_2$ makes a small contribution to the DOS near $\EF$. This is in line with the previously explained qualitative behavior of the self-energies. The left column of Fig.~\ref{fig:surfdos2} [panels (b), (d) and (f)] shows how the surface DOS evolves in the neighborhood of the adatom. Vertically above the adatom [Fig.~\ref{fig:surfdos2}(b)] only the $A_1$ orbital contribution is seen, as expected from symmetry considerations. Moving sideways reveals the contribution from the $E_1$ orbital [Fig.~\ref{fig:surfdos2}(d)], with a different Fano profile in comparison to the $A_1$ orbital. Considering a position at the same height above the surface as the adatom [Fig.~\ref{fig:surfdos2}(f)] shows that the Fano profiles are not only distance-dependent but also orientation and orbital-dependent. As the distance to the adatom increases, the Fano profile becomes dominated by the $A_1$ contribution, as can be seen in the right column of Fig.~\ref{fig:surfdos2} [panels (c), (e) and (g)]. This is probably due to the strong coupling between the $A_1$ orbital and the Cu(111) Shockley surface state, leading to long-range RKKY oscillations. The silent role of the $E_2$ orbital in the Fano profiles away from the adatom is due to its qualitatively different self-energy, that does not lead to a strong quasiparticle peak at $\EF$, and thus fails to couple strongly to the itinerant surface electrons.

The results discussed above demonstrate that STM spectra of adatoms indeed exhibit Fano line shapes, as predicted qualitatively in Ref.~\onlinecite{Ujsaghy2000} for a single orbital model. However, the multi-orbital character of the 3$d$ shell of the adatoms must be taken into account, as different orbitals possess qualitatively different single-particle hybridization functions and self-energies~\cite{Surer12,Khajetoorians2015,Frank2015}. Furthermore, the STM spectrum in the neighborhood of the adatom cannot be analyzed solely in terms of the symmetries of the adatom 3$d$ orbitals, as they couple with different strengths to the surrounding surface conduction electrons. It is the combination of all these effects which ultimately determines the spatial variation of the local density of states that is detected in tunneling experiments.

\section{Conclusions\label{sec:conclusions}}

In this work, we have investigated STM spectra of Co and Mn adatoms on the $(111)$ surface of copper within the Tersoff-Hamann approach, focusing on the role of electronic correlations within the $d$ shell of the adsorbed atoms. We used a combination of first-principles DFT calculations, based on the KKR approach, with CT-QMC and ED impurity solvers to evaluate the local density of states as a function of tip position. The dependence of the local electronic correlations on several parameters, such as onsite interaction $U$, temperature $T$ and filling $N_d$, was studied in detail in order to analyze the variation of the Kondo temperature or the coherence scale with these system parameters.

Two opposite limits, the Mn adatom in the local moment regime, and the Co adatom showing Kondo-like behavior, were chosen to illustrate different types of STM spectra. The Fano line shapes obtained in the Co adatom case were analyzed as a function of distance and the contributions from each correlated orbital. The results provide clear evidence for the importance of multi-orbital effects resulting both from the single-particle coupling to the substrate and the Coulomb correlations within the adatom. Thus, experimental STM spectra of transition mental adatoms can not be adequately analyzed in terms of single-orbital models. 

The key ingredient for the connection between the local correlated orbitals and the rest of the system is the hybridization function. Our calculations for Mn and Co reveal that, for the Cu(111) surface, this quantity is not very sensitive to the chemical identity of the 3$d$ adatom. Thus, from a model point of view, the hybridization function can be kept fixed, with different adatoms represented by different filling levels $N_d$.

Within the general multiorbital impurity model [Eq.~\eqref{eq:impurity_model}], we have investigated further the Kondo physics including both the spin fluctuation (as in the traditional Kondo model) and the charge fluctuation, where $N_d$ is varied in a wide range. The robust tendency is that, due to the charge fluctuation, the Kondo scale (or the coherence scale) is enhanced when the filling is away from integer, while at integer filling, the charge fluctuation is suppressed. Also, the Kondo temperature depends strongly on the magnitude of the local moment. Therefore, at half filling, when the local moment is largest and thus more likely to be frozen, the Kondo effect is suppressed, while at other integer filling, the local moment is smaller, allowing for a higher Kondo scale. For noninteger filling, the charge fluctuation becomes important and is the main reason for the increase of $T_K$. It is then not possible to work with the low-energy Kondo model, and therefore a general interaction such as Eq.~\eqref{eq:impurity_model} is required to study this problem.

The effect of correlation strength (represented by $U$) also depends on the $d$ occupancy when $U$ is in the range of interest ($4\to 5$~eV). Away from integer filling, as long as $N_d$ is adjusted to be the same, the physics for systems with different $U$ is similar. The effect of increasing $U$ is, however, clearly seen for integer filling, especially in the half-filling case where the correlation effect is largest. These findings again emphasize the important role of $N_d$ in controlling the physics of the system. On the other hand, the temperature variation mainly governs the distance from the Kondo regime and slightly induces orbital ordering for systems far from half-filling.

There are certain limitations in our study. Firstly, although the CT-QMC impurity solver \cite{Werner06} can handle general interactions, the calculations are restricted to high temperatures ($T\ge 0.0125$~eV). To investigate lower temperatures, we have also used ED as an impurity solver, which has the advantage that it becomes computationally less demanding at lower $T$. ED results for Co adatoms down to $T=0.0025$~eV indicate that at this temperature the system is still above the Kondo temperature, in particular, for the more strongly correlated orbitals. Qualitatively this result is consistent with experiment. We note, however, that because of the small bath size for $d$ electron impurities (at present two bath levels per orbital, 15 levels in total), the projection of the non-interacting adatom Green's function onto a small cluster becomes progressively less accurate at low $T$. Thus, using ED as well as CT-QMC, the orbital-dependent Kondo temperatures $T_{K,m}$ can at present only be estimated via an extrapolation of the adatom self-energy derived at high temperatures.

Second, the combination of DFT and impurity solvers poses the double-counting issue, as electronic correlations are included in both schemes and cannot easily be separated. A closely related problem is the determination of $N_d$ appropriate for each kind of adatom. As the physics is controlled mostly by $N_d$ it has to be determined consistently and from first-principles. A self-consistent treatment of the coupling between the correlated adatom orbitals and the surface electronic structure would answer this problem, provided a suitable approach to the double-counting is available \cite{Haule2014,Park2014a}.

Lastly, from the STM point of view, a more realistic description of the tunneling could be adopted. However, this step faces the same unknown as its experimental counterpart: the structure and composition of the STM tip. In the present approach, the averaging volume chosen for computing the surface DOS away from the adatom corresponds to roughly one atom, and $s$-like tunneling. How the picture changes if the tunneling is dominated by a tip orbital of different symmetry is left to further work \cite{Palotas2012}.

\section*{Acknowledgments}
H.~T.~D thanks A.~J.~Millis for valuable discussions. We acknowledge the allocation of computing time at J\"ulich Supercomputing Centre and RWTH Aachen University through JARA-HPC. H.~T.~D. acknowledges support from the Deutsche Forschungsgemeinschaft (DFG) within projects FOR 1807 and RTG 1995. M.~d.~S.~D and S.~L. are supported by the HGF-YIG Programme VH-NG-717 (Functional Nanoscale Structure and Probe Simulation Laboratory -- Funsilab).

\appendix

\section{Exact diagonalization}\label{app:exactdiagonalization}

According to the results shown in Fig.~\ref{fig:TkSplot}, the orbital dependent Kondo temperatures evaluated from Eq.~\eqref{eq:kondoT} for Co adatoms on Cu(111) lie in the range $T_K\approx 0.01\to 0.04$~eV. Thus, they are of the same order or smaller than the temperatures which we have used within the present CT-HYB calculations ($T\ge0.0125$~eV). Calculations at lower temperatures become increasingly computationally expensive. During the recent years it was shown that the correlated impurity problem of five-orbital systems can also be solved within exact-diagonalization (ED) method \cite{Liebsch2012}. For instance, it has been demonstrated that compounds such as LaFeAsO and FeSe exhibit a spin-freezing transition as a function of doping \cite{Ishida2010,Liebsch2010}.

\begin{figure}[h]
\centering
 \includegraphics[width=0.8\columnwidth]{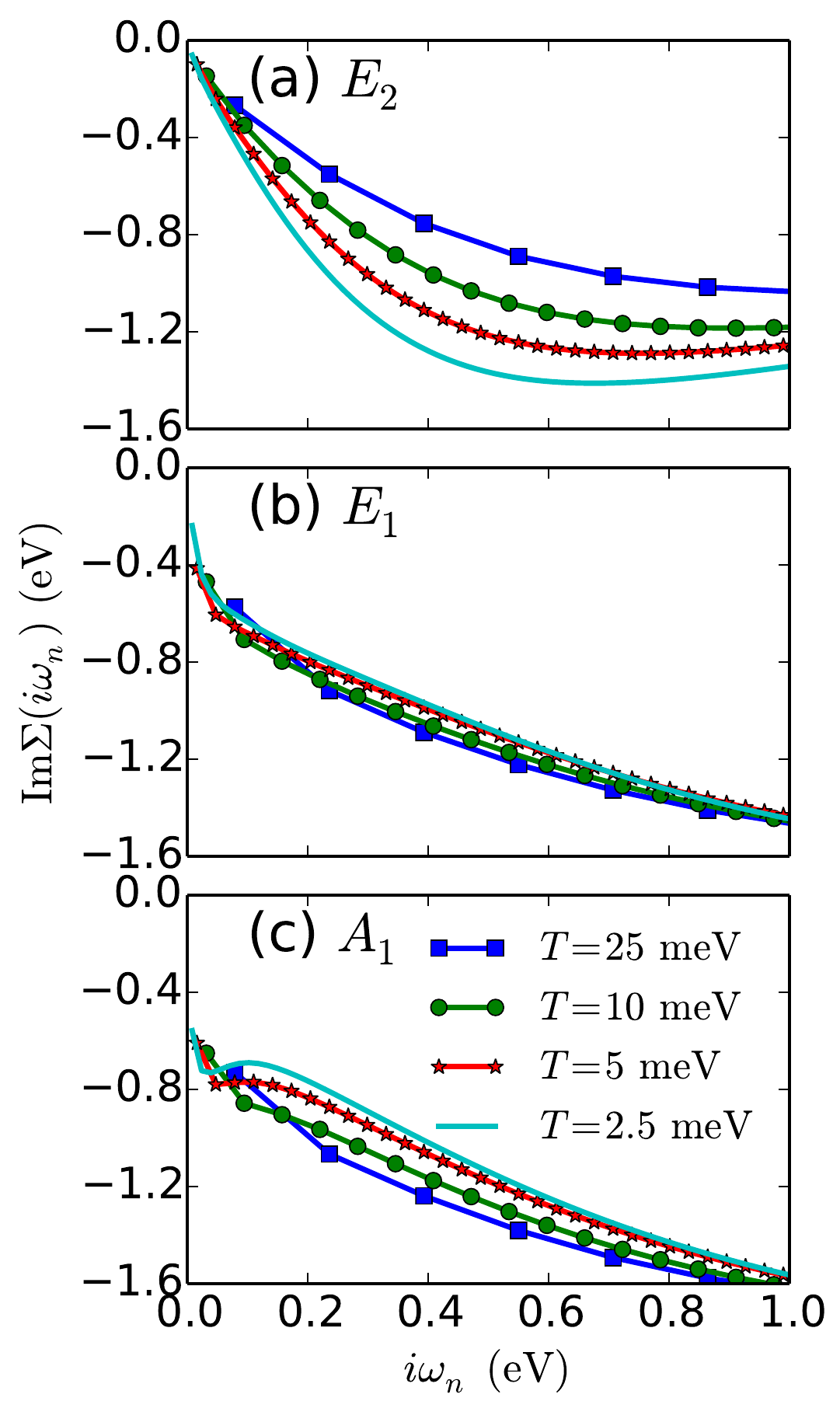}
\caption{\label{fig:ED}(Color online) Orbital components of imaginary part of self-energy for Co on Cu(111), evaluated within ED at temperatures $T=0.0025,\ 0.005,\ 0.01,\ 0.04$~eV at $\mu=26$~eV, $U=4$~eV, and $J=0.9$~eV.}
\end{figure}

ED may be viewed as complementary to QMC in the sense that it becomes computationally less costly at lower temperatures as exponentially fewer excited states must be included in the evaluation of the impurity Green's function. In fact, it is possible to only consider the ground state, i.e., to investigate the $T\rightarrow 0$ limit. However, approaching this limit the ED impurity calculation becomes progressively less accurate since the non-interacting Green's function $G^0_m(i\omega_n)$ [see Eq.~\eqref{eq:gfmatsu} for $\Sigma_m(i\omega_n)=0$] must first be fitted to a Green's function corresponding to an impurity orbital immersed in a bath with a finite number of energy levels. Thus, effectively the continuum of conduction states hybridizing with the impurity is discretized via a finite cluster. In a single-orbital system, highly accurate fits can readily be achieved using $8$ to $12$ cluster levels, so that temperatures down to $T=0.001$~eV or less can be considered \cite{Ishida2012}.

In a multi-orbital system, however, the number of available bath levels per orbital rapidly decreases. As the overall cluster size at present is limited to about $15$, for a $d$ shell only two bath levels per orbital are feasible. As a result, ED calculations at temperatures less than about $0.01$~eV become progressively less accurate.

To illustrate this point, we show in Fig.~\ref{fig:ED} the self-energy components for the Co adatom at several temperatures. For simplicity, the Lorentzian density of states profiles are used as input, where the energy levels and line widths are specified in Table~\ref{table:kkr_dos}. For all orbitals, we find a clear trend, namely, that the initial slope of the self-energy increases at lower $T$, indicating that the system has not yet reached the Fermi-liquid regime. The Kondo temperatures derived from Eq.~\eqref{eq:kondoT} therefore diminish with $T$, according to the decreasing values of $Z_m$. Although this trend is expected in view of the experimentally observed small values of $T_K$, the kinks appearing in the self-energy at the lowest Matsubara frequencies indicate that inaccuracies resulting from the small cluster size become more pronounced, in particular, for the more correlated orbitals $A_1$ and $E_1$. It therefore is not possible to extract reliable quasi-particle weights at very low $T$.

\section{Lorentzian fitting \label{app:lorentzian}}

The shapes of the input DOS obtained from KKR calculations shown in Fig.~\ref{fig:kkr_dos} are very similar to Lorentzian distributions [Eq.~\eqref{eq:lorentzian}]. Therefore, we investigate how the results are modified if the true DOS spectra are approximated by Lorentzian DOS profiles. 

\begin{figure}[t]
\centering
 \includegraphics[width=\columnwidth]{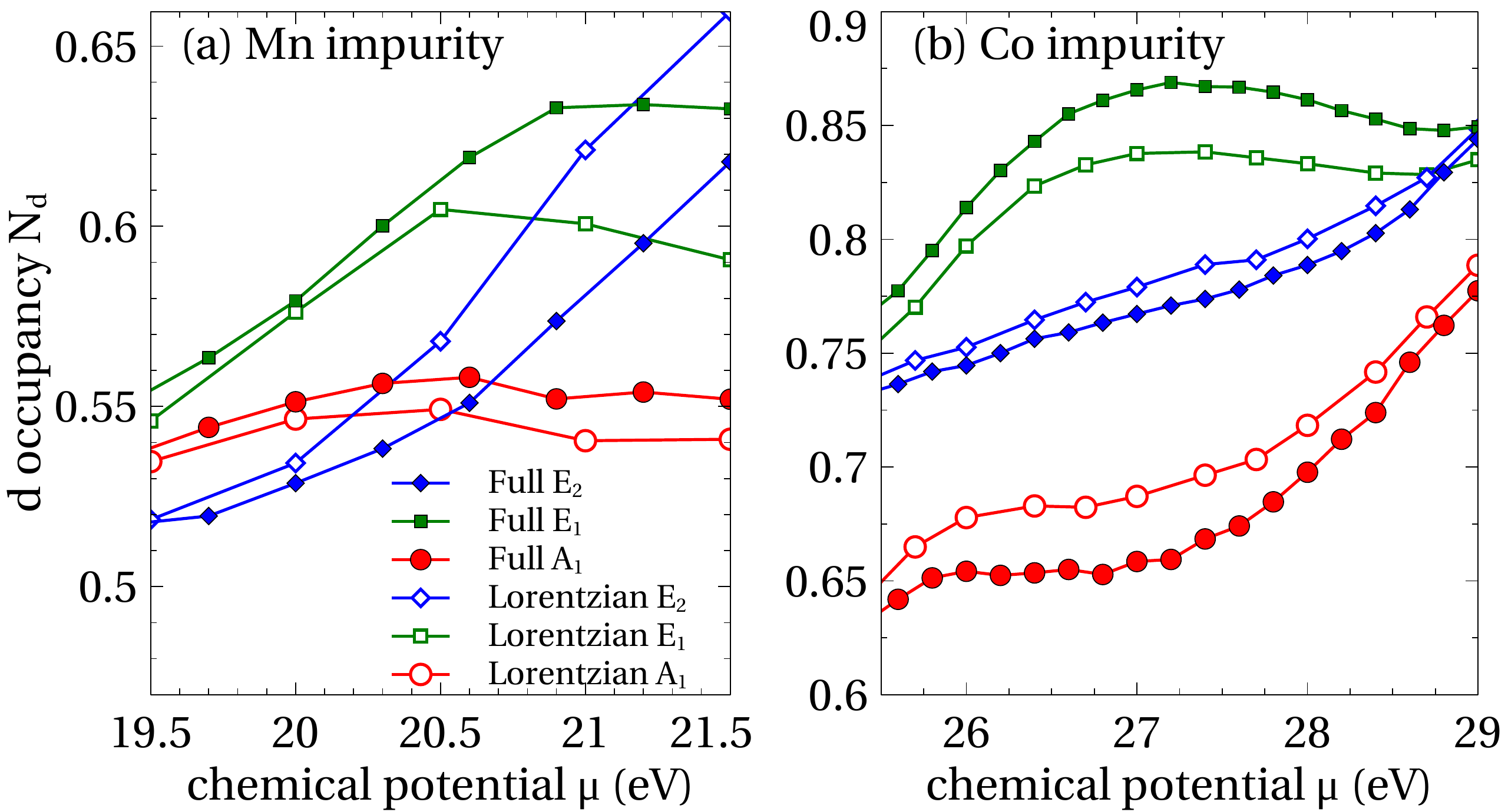} 
\caption{\label{fig:lorentzian_occupancy}(Color online)  Comparison of chemical potential dependent orbital occupancies of Mn (a) and Co (b) adatoms, obtained from a calculation using input KKR (``full'') DOS or Lorentzian DOS. Open symbols are for Lorentzian DOS, closed symbols are for full DOS. Both use the same parameters $U=4$~eV, $J=0.9$~eV, and $T=0.05$~eV.}
\end{figure}

The Lorentzian form is obtained by fitting the noninteracting DOS within a narrow range of energy near the Fermi level using Eq.~\eqref{eq:lorentzian}. The energy window from $-1.45$~eV to $4$~eV is chosen so that it includes only the Lorentzian DOS peaks [see Fig.~\ref{fig:kkr_dos}] or, equivalently, the flat area of the hybridization functions [see Fig.~\ref{fig:hyb}]. The fitting parameters are summarized in Table~\ref{table:kkr_dos} in comparison with $\epsilon_m$ and $\Gamma_m$ for the actual DOS distributions. In the region of the Cu $3d$ bands around $-3$~eV [see Figs.~\ref{fig:kkr_dos} and \ref{fig:hyb}], $E_2$ orbitals have the smallest contribution, followed by $A_1$, $E_1$ orbitals have the largest weight. It thus reflects the changes from $\epsilon_m$ to $\epsilon^L_m$ for each orbital.

To understand how these changes affect the final results with interaction, we carry out the investigation by running the same CT-QMC simulations but with the Lorentzian DOS input at $U=4,J=0.9$~eV, $T=0.05$~eV and the chemical potential $\mu$ varied in a wide range. The results are then compared with those using the full noninteracting DOS as input.

\begin{figure}[t]
\centering
\includegraphics[width=\columnwidth]{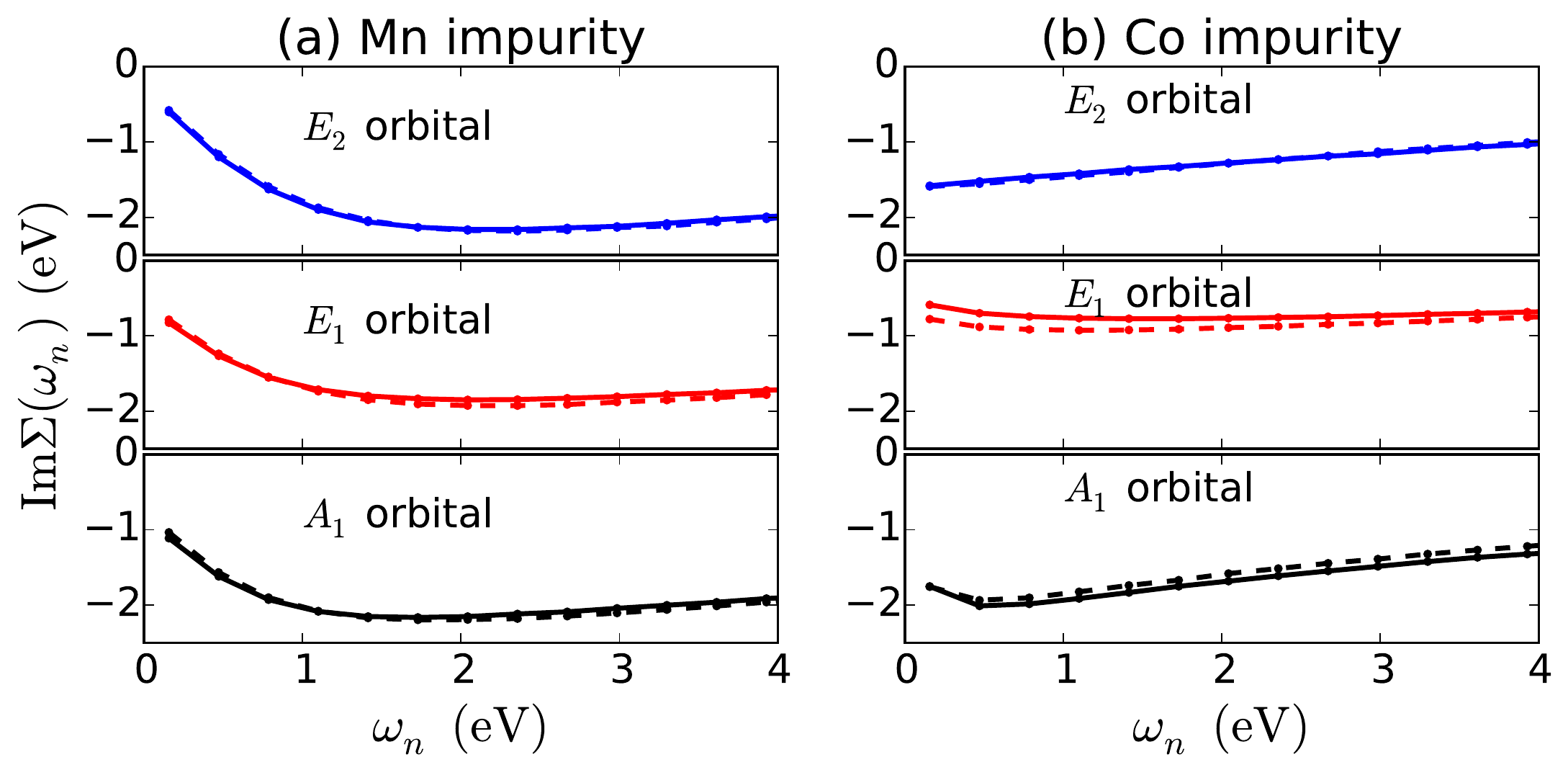}
\caption{\label{fig:lorentzian_selfenergy}(Color online) Comparison of imaginary part of the Matsubara self-energy for KKR (``full'') DOS (solid curves) and Lorentzian DOS (dashed curves) as a function of chemical potential; $U=4eV,J=0.9$~eV and $T=0.05$~eV. The chemical potential is the same for both input KKR and Lorentzian DOS: $\mu=20$~eV for Mn, $\mu=27$~eV for Co.}
\end{figure}

Figure~\ref{fig:lorentzian_occupancy} shows the comparison of the orbital occupancies using the two types of input DOS. While the total $N_d$ and the total spin $S$ are similar for the two cases (not shown), the orbital occupancies exhibit certain differences. $E_1$ orbital has the largest DOS portion at $-3$~eV, in the Lorentzian form, this DOS part is neglected, $E_1$ occupancy thus decreases. In contrast, $E_2$ orbital is the most localized of the three orbitals, in the Lorentzian form almost all of its DOS is included, its occupancy is increased. $A_1$ orbital is in the intermediate. Moreover there is only one orbital $3z^2-r^2$ in the $A_1$ group, while there are two orbitals in $E_2$ or $E_1$ groups, the tendency of $A_1$ occupancy is more likely to change. Thus its occupancy is increased in the Co case [Fig.~\ref{fig:lorentzian_occupancy}(b)] and decreased in the Mn case [Fig.~\ref{fig:lorentzian_occupancy}(a)].

Interestingly, the self-energy does not exhibit significant change, as shown in Fig.~\ref{fig:lorentzian_selfenergy} for the imaginary part of the Matsubara self-energy. Normally, the DOS part far from the Fermi level has the effect of screening the on-site interaction \cite{Dang14,Haule2014}, however, in Fig.~\ref{fig:lorentzian_selfenergy}, only the $E_1$ orbital becomes slightly more correlated in the Lorentzian input DOS, other orbitals show nearly the same self-energy. We have checked with the Kanamori interaction (not shown), which exhibits larger increase in the correlation strength for Lorentzian input DOS. It suggests that the results from rotationally invariant interaction is more stable against small perturbation.

Therefore, by using the Lorentzian DOS obtained from fitting the KKR DOS as input for the impurity solver, our main results are not much affected, only the orbital occupancies are changed due to the disappearance of the DOS part far from the Fermi level, mostly composed of Cu $3d$ character around $-3$~eV. The comparison suggests that Lorentzian fitting is a good approximation for the study of the impurity problem. It may allow for the construction of a simple model to study a wide range of $3d$ impurity systems.

\bibliographystyle{apsrev4-1}
\bibliography{surface_impurity}

\end{document}